\documentclass[12pt]{article}
\usepackage{eurosym}
\usepackage[left=0.0cm, right=0.0cm, top=0.1in, bottom=0.1in]{geometry}
\usepackage{hyperref}
\usepackage{mathtools}
\usepackage{graphicx}
\usepackage{mathrsfs}
\usepackage{cite}
\newcommand{\bra}[1]{\bigl\langle #1 \bigr|}
\newcommand{\ket}[1]{\bigl| #1 \bigr\rangle}
\begin{document}
	\begin{center}
\textbf{\Large Improving the  bidirectional steerability  between two accelerated partners via filtering process\\}
	\bigskip	
M.Y.Abd-Rabbou$^{a}$ \textit{\footnote{e-mail: m.elmalky@azhar.edu.eg}},
 N. Metwally$^{b,c}$\textit{\footnote{e-mail:nmetwally@gmail.com}}, M. M. A. Ahmed $^{a}$,  and A.-S. F. Obada $^{a}$

$^{a}${\footnotesize Mathematics Department, Faculty of Science,
Al-Azhar University, Nasr City 11884, Cairo, Egypt}\\
$^{b}${\footnotesize Math. Dept., College of Science, University of Bahrain, Bahrain.}\\

$^{c}${\footnotesize Department of Mathematics, Aswan University
	Aswan, Sahari 81528, Egypt.}\\
	\end{center}

\begin{abstract}

The bidirectional steering between two accelerated partners sharing initially different classes of entangled states is discussed. Due to the decoherence, the steerability and its degree decrease either as the acceleration increases or the partners share initially a small amount of quantum correlations. The possibility of increasing the steerability is investigated by applying the filtering process. Our results show that by increasing the filtering strength, one can improve the upper bounds of the steerability and the range of acceleration at which the steerability is possible. Steering large coherent states is much better than steering less coherent ones.
	
	\end{abstract}


\section{Introduction}
The concept of Einstein–Podolsky–Rosen (EPR) steering is suggested by Schr{\"o}dinger in 1935 to discuss the EPR paradox \cite{1,2}. EPR steering is a significant quantum phenomenon in which a quantum state can be non-locally changed or steered another state remotely by performing some local measurements \cite{ddd}. Besides, the violation of EPR steering violation has been set as a measure of quantum correlations, which is laid out between the non-separability and Bell non-locality \cite{3,4}. Several experimental studies deal with interpreting the dynamics of quantum steering, such as; EPR steering game employing an all-versus-nothing criterion that depends on obtaining two different pure normalized conditional states has been explored \cite{5}. Measurement-device-independent steering protocols for optical polarization qubits and states that do not violate Bell inequality have been demonstrated \cite{exp1}.  By polarizing the photons in a linear-optical setup, the temporal quantum steering has been applied for testing and securing the quantum communications \cite{exp2}. The convex steering monotone and measure of steerability depend on the observed statistics and the quantum inputs have been discovered \cite{7}. Quantum steering non-locality of high-dimensional quantum systems with isotropic noise fraction has been retrieved \cite{9}. Theoretically, the steering inequality and steerability have been reconstructed by using different relations as Heisenberg uncertainty principle \cite{11}, the standard geometric Bell inequalities \cite{12,13}, a version of the Clauser-Horne-Shimony-Holt inequality \cite{14,15},  fine-grained uncertainty relation\cite{16},  Abner Shimony inequalities \cite{jzhao}, and steering witnesses\cite{17}. The violating  of EPR inequalities indicates the possibility of steering process  \cite{18,19}.

 Due to the importance of steering in determining the quantum correlation, it has been used in many practical applications of quantum informatics. For example, it has been applied to investigate the security and feasibility of a one-sided device of  the standard quantum key distribution \cite{20}. It also has been employed in quantum computation \cite{21} and securing quantum teleportation \cite{22}. Moreover, quantum steering has been  discussed for some quantum systems such as, bipartite two-qubit X-state \cite{23,24}, Heisenberg chain models \cite{25}, two-level or three-level detectors \cite{26}.  As quantum systems inevitably interact with the different environments, it is  necessary to discuss the influence of these environments on the steering process. For instance, the effect of asymmetric dissipation relativistic motion \cite{27}, finite temperature \cite{30}, non-Markovian environment on the steering \cite{31}, and noise channel \cite{33} have been discussed. The steering entropic uncertainty of the qutrit system under amplitude damping decoherence has been examined \cite{Huang}.
 
 In addition, the possibility of improving of quantum correlations via local filtering operation has been widely investigated, such as, improving the steering and nonlocality in Heisenberg XY mode \cite{Zhao}, also protecting entanglement from the decoherence due to amplitude damping and acceleration\cite{zz11,nn11}. On the other hand, investigating the quantum correlation in an  accelerated frame is one of the considerable areas in theoretical quantum processing, where the actual quantum systems are essentially non-inertial \cite{34,35}. The acceleration process may cause  a dissipation of quantum correlations between an inertial observer and an accelerated one \cite{36}. Under the non-inertial frame, the  entanglement of  a general two-qubit system has been studied \cite{37}. The classicality and quantumness under different decoherence noisy channels in accelerated frame were investigated in \cite{38}. The general behaviour of EPR steering inequality and steerability between two users have been investigated, where the first in flat space-time and another in the non-inertial frame \cite{39}.

Our motivation in this work is to derive a generic version of the steering inequality for two quantum  subsystems with two dimensions (two-qubit). We then apply the new inequality on a general form of a two-qubit system, as it is possible to generate some special classes of them \cite{40}. We assume that the actual quantum regimes are essentially in the non-inertial frame, where the relativistic framework is taken into account, whether we accelerate only one subsystem or accelerate the two subsystems collectively.

The paper is organized as follows: In Sec.\ref{S1}, we derive a general form of steering inequality for a bipartite qubit system. In Sec. (\ref{sf}) we are defined the filtering process of the two-qubit state to improve the efficiency of shared entangled state. In Sec. (\ref{S2})  the EPR steering under the acceleration process of the generalized Werner state and the generic pure state has proposed. Additionally, the degree of steerability is enhanced  via using the local filtering operation. Bidirectional steerability for the  generic pure state from Alice to Bob is studied in Sec.\ref{S3}, where Alice's particle is accelerated and Bob's particle is  at rest. Finally, we conclude our results in Sec.(\ref{s6}).

\section{Entropic Steering:}\label{S1}
In this section, we derive a generalized form of quantum steering for a bipartite system. According to the definition in \cite{11}, the entropic uncertainty steering inequality for the discrete observable in the even N-dimensional Hilbert space is defined as,
\begin{equation}
	\sum_{i=1}^{N+1}H(R_{b_i}|R_{a_i})\geq \frac{N}{2} Log_2(\frac{N}{2})+(1+\frac{N}{2}) Log_2(1+\frac{N}{2}),
\end{equation}
where $H(R_b|R_a)=H(\rho_{ab})- H(\rho_{a})$ is the conditional Shannon entropy for the  subsystems $a$ and $b$. By applying the Pauli spin  operator ($\sigma_x,\sigma_y,\sigma_z$) as measurements, the EPR inequality of steering from  A to B is given by,
\begin{equation} \label{e1}
\mathcal{I}_{ab}=H(\sigma_x^{(b)}|\sigma_x^{(a)})+H(\sigma_y^{(b)}|\sigma_y^{(a)})+H(\sigma_z^{(b)}|\sigma_z^{(a)}\geq 2.
\end{equation}

In the computational basis $|00\rangle,\ |01\rangle,\ |10\rangle $, and $|11\rangle$, the general form of an arbitrary two-qubit system may be written as:
\begin{equation}\label{e2}
\hat{\rho}_{ab}= \begin{pmatrix}
\rho_{11} & \rho_{12} & \rho_{13}& \rho_{14}\\\rho_{12}^*  & \rho_{22} & \rho_{23} & \rho_{24} \\ \rho_{13}^*  & \rho_{23}^* & \rho_{33} & \rho_{34}\\ \rho_{14}^* & \rho_{24}^* & \rho_{34}^*& \rho_{44} \end{pmatrix},
\end{equation}
By using Pauli matrices $\sigma_x$, $\sigma_y$, and $\sigma_z$ measurements, one  can simplify the expression of EPR steering inequality in Eq.(\ref{e1}) reads,
\begin{equation}\label{e3}
	\begin{split}
	\mathcal{I}_{ab}=&\frac{1}{2}\sum_{i=1}^{4}\bigg\{P^{ab}_{x_i}\log_2[P^{ab}_{x_i}]+P^{ab}_{y_i}\log_2[P^{ab}_{y_i}]+P^{ab}_{z_i}\log_2[P^{ab}_{z_i}]\bigg\}
\\&-\sum_{i=1}^{2}\bigg\{P^{a}_{x_i}\log_2[P^{a}_{x_i}]+P^{a}_{y_i}\log_2[P^{a}_{y_i}]+P^{a}_{z_i}\log_2[P^{a}_{z_i}]\bigg\}
	\end{split},
\end{equation}
where $P^{ab}_{x_i},\ P^{ab}_{y_i}$, and $P^{ab}_{z_i}$ represent the  eigenvalues of the density operator $\hat{\rho}_{ab} $. Explicitly,  they are given by
\begin{equation}
	\begin{split}
	&P^{ab}_{x_1}=(1+2\ Re[\rho_{12}+\rho_{13}+\rho_{14}+\rho_{23}+\rho_{24}+\rho_{34}]),\\&
	P^{ab}_{x_2}=(1-2\ Re[\rho_{12}+\rho_{13}-\rho_{14}-\rho_{23}+\rho_{24}+\rho_{34}]),\\&
	P^{ab}_{x_3}=(1-2\ Re[\rho_{12}-\rho_{13}+\rho_{14}+\rho_{23}-\rho_{24}+\rho_{34}]),\\&
	P^{ab}_{x_4}=(1+2\ Re[\rho_{12}-\rho_{13}-\rho_{14}-\rho_{23}-\rho_{24}+\rho_{34}]),\\&
	P^{ab}_{y_{1},y_{2}}=(1+2\ Re[\rho_{23}-\rho_{14}]\pm 2\ Im[\rho_{12}+\rho_{13}+\rho_{24}+\rho_{34}]),\\&
	P^{ab}_{y_{3},y_{4}}=(1-2\ Re[\rho_{23}-\rho_{14}]\pm 2\ Im[\rho_{12}-\rho_{13}-\rho_{24}+\rho_{34}]),\\&
	P^{ab}_{z_i}=4 \rho_{ii}.
	\end{split}
\end{equation}
 Likewise, $P^{a}_{x_i},\ P^{a}_{y_i}$, and $P^{a}_{z_i}$ are the  eigenvalues of the reduced density operator $\hat{\rho}_{A} $, where,
\begin{equation}
	\begin{split}
	&P^{a}_{x_1,x_2}=(1\pm 2\ Re(\rho_{13}+\rho_{24})),\quad P^{a}_{y_1,y_2}=(1\pm 2\ Im(\rho_{13}+\rho_{24})), \\&
	 \text{and} \quad P^{a}_{z_1,z_2}=(1\pm(\rho_{11}+\rho_{22}-\rho_{33}-\rho_{44})).
	\end{split}
\end{equation}
The  degree of steerability  is quantified based on Alice’s measurements as follows \cite{21},
\begin{equation}
	\mathcal{S}^{A\longrightarrow B}=\max\bigg\{0,\frac{\mathcal{I}_{ab}-2}{\mathcal{I}_{max}-2}\bigg\},
\end{equation}
where $\mathcal{I}_{max}$=6 is calculated  for a  system  initially prepared in Bell states. The factor $( \mathcal{I}_{max}-2)$ is introduced to ensure that the steering  process is normalized. By exchanging the roles of A and B, the possibility of the steering by performing measurements on the subsystem B is  given by,
\begin{equation}
\mathcal{S}^{B\longrightarrow A}=\max\bigg\{0,\frac{I_{ba}-2}{I_{max}-2}\bigg\},
\end{equation}
where $\mathcal{I}_{ba}$ quantifies the steering  from Bob  to Alice, it is given by,
\begin{eqnarray}\label{e9}
\mathcal{I}_{ba}&=&\frac{1}{2}\sum_{i=1}^{4}\bigg\{P^{ab}_{x_i}\log_2[P^{ab}_{x_i}]+P^{ab}_{y_i}\log_2[P^{ab}_{y_i}]+P^{ab}_{z_i}\log_2[P^{ab}_{z_i}]\bigg\}
\nonumber\\
&-&\sum_{i=1}^{2}\bigg\{P^{b}_{x_i}\log_2[P^{b}_{x_i}]+P^{b}_{y_i}\log_2[P^{b}_{y_i}]+P^{b}_{z_i}\log_2[P^{b}_{z_i}]\bigg\},
\end{eqnarray}
with,
\begin{eqnarray}
P^{b}_{x_1,x_2}&=&(1\pm 2\ Re(\rho_{12}+\rho_{34})),\quad P^{b}_{y_1,y_2}=(1\pm 2\ Im(\rho_{12}+\rho_{34})),
\nonumber\\
 \quad P^{b}_{z_1,z_2}&=&(1\pm(\rho_{11}-\rho_{22}+\rho_{33}-\rho_{44})).
\end{eqnarray}
 Hereinafter, we shall discuss the Unruh effect on  the steering process of the system $\hat{\rho_{ab}}$  and the possibility of improving this process via  local filter operations.
\section{Filter Process:}\label{sf}
Filtering process is a method that may be used to  improve the efficiency of a shared entangled state between two users Alice and Bob to perform some quantum information tasks. As we shall see in Sec.4, the acceleration process decreases the efficiency of steerability, therefore the users  need to apply the filtering process on the accelerated shared state to improve its efficiency. The final output filtered  state $\hat{\rho}_{ab}^{\mathcal{F}}$  is given by \cite{29.1},
\begin{equation}
	\hat{\rho}_{ab}^{\mathcal{F}}=\frac{1}{\mathcal{N}}\mathcal{W}_a \mathcal{W}_b.\hat{\rho}_{ab}.\big(\mathcal{W}_a \mathcal{W}_b\big)^{\dagger},
\end{equation}
with $ \mathcal{N}= Tr[\mathcal{W}_a \mathcal{W}_b.\hat{\rho}_{ab}.\big(\mathcal{W}_a \mathcal{W}_b\big)^{\dagger}] $ and,
\begin{equation}
	\mathcal{W}_i=diag\{\sqrt{\alpha_i},\sqrt{1-\alpha_i}\}, \quad i=a\ \text{or} \ b.
\end{equation}
The parameters $ \alpha_i $ are the strengths of the filtering  process. After applying the operator $\mathcal{W}_i$, the final output state is given by,
\begin{equation}
	\hat{\rho}_{ab}^{\mathcal{F}}= \frac{1}{\mathcal{N}} \begin{pmatrix}
		\rho_{11} \alpha^2 & \rho_{12}  \sqrt{1-\alpha} \alpha^{\frac{3}{2}}& \rho_{13}   \sqrt{1-\alpha} \alpha^{\frac{3}{2}}& \rho_{14}  \alpha (1-\alpha)\\ \rho_{12}^*    \sqrt{1-\alpha} \alpha^{\frac{3}{2}} & \rho_{22} \alpha (1-\alpha) & \rho_{23} \alpha (1-\alpha)  & \rho_{24} \sqrt{\alpha (1-\alpha)^3}\\ \rho_{13}^*   \sqrt{1-\alpha} \alpha^{\frac{3}{2}} & \rho_{23}^* \alpha(1-\alpha) & \rho_{33} \alpha (1-\alpha) & \rho_{34} \sqrt{\alpha (1-\alpha)^3} \\ \rho_{14}^* \alpha (1-\alpha)& \rho_{24}^* \sqrt{\alpha(1-\alpha)^3}& \rho_{34}^* \sqrt{\alpha(1-\alpha)^3}& \rho_{44} (1-\alpha)^2 \end{pmatrix},
\end{equation}
where it is  assumed that, the strength parameters of the filtering  process are equal.

\section{EPR steering under the Unruh thermal framework:}\label{S2}

Let us consider that, the users Alice and Bob share a general two-qubit  system  on the following form,

 \begin{equation}\label{state}
 	\hat{\rho}_{ab}=\frac{1}{4}\bigg(I^{(a)}_2\otimes I^{(b)}_2+ \vec{s}.\sigma\otimes I^{(b)}_2+I^{(a)}_2\otimes \vec{t}.\tau+\sum_{i,j}^{3}c_{ij} \sigma_i\otimes\tau_j \bigg), i,j=x,y,z,
 \end{equation}
 where  $I_{2}^{a(b)}$  represents  the identity operators of  Alice's (a) and Bob's (b) qubits. The vectors $\vec{s}=(s_x~,s_y, ~s_z)$, and $\vec{t}=(t_x,t_y,t_z)$ are the Bloch vectors of Alice's and Bob's qubit, respectively.  The operators  $ \sigma_i, \tau_i, i=1..3$,  are the standard Pauli  spin operators, where $\sigma_x=\tau_x=\ket{0}\bra{1}+\ket{1}\bra{0}$, $\sigma_y=\tau_y=i(\ket{0}\bra{1}-\ket{1}\bra{0})$ and $\sigma_z=\tau_z=\ket{0}\bra{0}-\ket{1}\bra{1}$. The correlation matrix (dyadic), $c_{ij}$ is a $3\times 3$ matrix, where its elements are defined by $c_{ij}=Tr\{\sigma_i\tau_j\rho_{ab}\},~ i,j=1,2,3$.

Since our aim is to discuss the phenomena of the quantum steering by using an accelerated state between Alice and Bob, then it is important to  review Unruh effect on
Minkowski's particles. However, in the inertial frames, Minkowsik
coordinates $(t,x)$ are  used to describe the dynamics of
these particles, while in the non-inertial frames Rindler
coordinates $(\zeta, \chi)$ are used. The relations between the
Minkowski and Rindler's spaces are given by,
\begin{equation}\label{trans}
\zeta=r~tanh\left(\frac{t}{x}\right), \quad \chi=\sqrt{x^2-t^2},
\end{equation}
where  $-\infty<\zeta<\infty$, $-\infty<\chi<\infty$  and $r$ is
the acceleration of the moving particle. The transformation
(\ref{trans}) defines two regions in Rindler's space: the first
region $I $ for $|t|<x$  and the second region $II$ for $x<-|t|$.

 A single mode $k$ of fermions and anti-fermions in Minkowski space  is described by the
annihilation operators $a_k$  and $b_{-k}$ respectively, where
$a_k\ket{0_k}=0$  and $b^\dagger_{-k}\ket{0_{k}}=0$.
 Based on the  approximation by  Bruschi et. al
 \cite{34, 41},  the relations between Minkowski and Rindler's operators are given
by Bogoliubov transformation,
\begin{equation}\label{op}
a_k=\cos r c^{(I)}_k-\exp(-i\phi)\sin r d^{(II)}_{-k}, \quad
b^\dagger_{-k}=\exp(i\phi)\sin r c^{(I)}_k+\cos r d^{(II)}_{k},
\end{equation}
where $c^{(I)}_k, d^{(II)}_{-k}$ are the annihilation operators of
Rindler's space in the regions $I$ ( for fermions) and $II$ (for
anti-fermions) respectively, $tan r=e^{-\pi\omega \frac{c}{a}}$,
$0\leq r\leq \pi/$4, $ a$ is the acceleration such that $0\leq
a\leq\infty$, $\omega$ is the frequency of the travelling qubits,
$c$ is the speed of light and $\phi$ is an unimportant phase that
can be absorbed into the definition of the operators
\cite{42,43}. The transformation (\ref{op}) mixes a
particle in first  region $I$ and an anti particle in the second region $II$\cite{44}. In terms of
Rindler's modes, the  Minkowski vacuum $\ket{0_k}_M$ and the one
particle state $\ket{1_k}_M$ take the forms,
\begin{eqnarray}\label{vacuum1}
\ket{0_k}_M&=&\cos r\ket{0_k}_I\ket{0_{-k}}_{II}+ \sin
r\ket{1_k}_I\ket{1_{-k}}_{II}, \nonumber\\
\ket{1_k}_M&=&a^\dagger_k\ket{0_k}_M =\ket{1_k}_I\ket{0_k}_{II}.
\end{eqnarray}
Then, by using these transformations and after  tracing out the degree of freedom in the second region $ II $, the initial state  (\ref{e2}) may be written as,
\begin{equation}\label{racc}
\hat{\rho}_{ab}^{acc}=\begin{pmatrix}
 \mathcal{C}^2_a \mathcal{C}^2_b \rho_{11}& \mathcal{C}^2_a \mathcal{C}_b \rho_{12} &\mathcal{C}_a \mathcal{C}^2_b \rho_{13}& \mathcal{C}_a \mathcal{C}_b \rho_{14}\\
\mathcal{C}^2_a \mathcal{C}_b \rho_{12}^* &\mathcal{C}^2_a \big(\rho_{22}+ S^2_b \rho_{11}\big) &\mathcal{C}_a \mathcal{C}_b \rho_{23}&\mathcal{C}_a \big(\rho_{24}+ S^2_b \rho_{13}\big)\\
\mathcal{C}_a \mathcal{C}^2_b \rho_{13}^*&\mathcal{C}_a \mathcal{C}_b \rho_{23}^*&\mathcal{C}^2_b \big(\rho_{33}+ S^2_a \rho_{11}\big) &\mathcal{C}_b \big(\rho_{34}+ S^2_a \rho_{12}\big)\\
\mathcal{C}_a \mathcal{C}_b \rho_{14}^*&\mathcal{C}_a \big(\rho_{24}^*+ S^2_b \rho_{13}^*\big)&\mathcal{C}_b \big(\rho_{34}^*+ S^2_a \rho_{12}^*\big)&S^2_a \big(\rho_{22}+ S^2_b \rho_{11}\big)+ S^2_b \rho_{33}+\rho_{44}
\end{pmatrix},
\end{equation}
 where $\mathcal{C}_i=\cos r_i$, and $S_i=\sin r_i, i=a,b $.

Now, we are ready to  investigate the steerability process between  Alice and Bob who share the accelerated state. This idea will be clarified by  using two different initial states namely; a generalized Werner state and a generic pure state.

\subsection{Generalized Werner state:}
This state is obtained by setting  $\vec{r}=\vec{0}=\vec{s}$, and the cross dyadic $C$ is defined by a  diagonalize  $3\times 3$ matrix, where $\{c_{ij}\}=diag\{c_{11},c_{22},c_{33}\}$. Then, Eq. (\ref{state}) reduces to be;
\begin{equation}\label{gw}
	\hat{\rho}_{ab}=\frac{1}{4}\bigg(I^{(a)}_2\otimes I^{(b)}_2+\sum_{i,j}^{3}c_{ij} \sigma_i\otimes\tau_j \bigg), i, j=x,y,z.
\end{equation}
Hence, the output accelerated system according to Eq.(\ref{racc}) is given by,
\begin{equation}
	\begin{split}
	\hat{\rho}_{ab}^{acc}= &\mathcal{C}^2_a \mathcal{C}^2_b\mathcal{A}_{11}|00\rangle \langle 00|+ \mathcal{C}^2_a \big(\mathcal{A}_{22}+ S^2_b \mathcal{A}_{11}\big) |01\rangle \langle 01|\\+&
	\mathcal{C}^2_b \big(\mathcal{A}_{22}+ S^2_a \mathcal{A}_{11}\big)|10\rangle \langle 10|\\+& (S^2_a \big(\mathcal{A}_{22}+ S^2_b \mathcal{A}_{11}\big)+ S^2_b \mathcal{A}_{22}+\mathcal{A}_{11})|11\rangle \langle 11|\\+& \mathcal{C}_a \mathcal{C}_b \big\{\mathcal{A}_{14}|00\rangle \langle 11|+\mathcal{A}_{23}|01\rangle \langle 10| +h.c.\big\},
	\end{split}
\end{equation}
where,
\begin{equation}
	\mathcal{A}_{11}=\frac{1+c_{33}}{4}, \quad 	\mathcal{A}_{22}=\frac{1-c_{33}}{4}, \quad 	\mathcal{A}_{14}=\frac{c_{11}+c_{22}}{4}, \quad 	\mathcal{A}_{23}=\frac{c_{11}-c_{22}}{4},
\end{equation}

\begin{figure}[h!]
	\centering
	\includegraphics[width=0.32\linewidth, height=3.7cm]{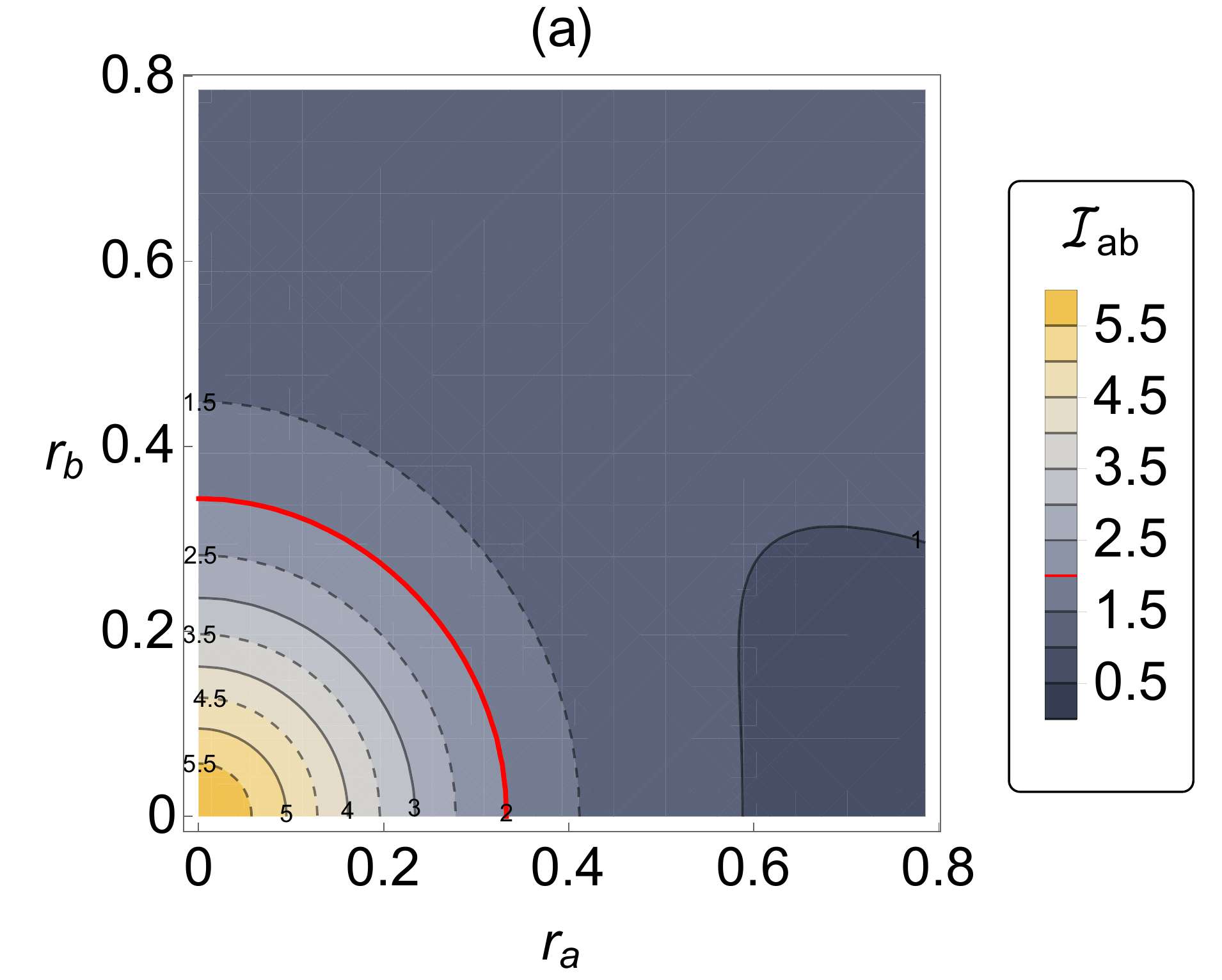}
	\includegraphics[width=0.32\linewidth, height=3.7cm]{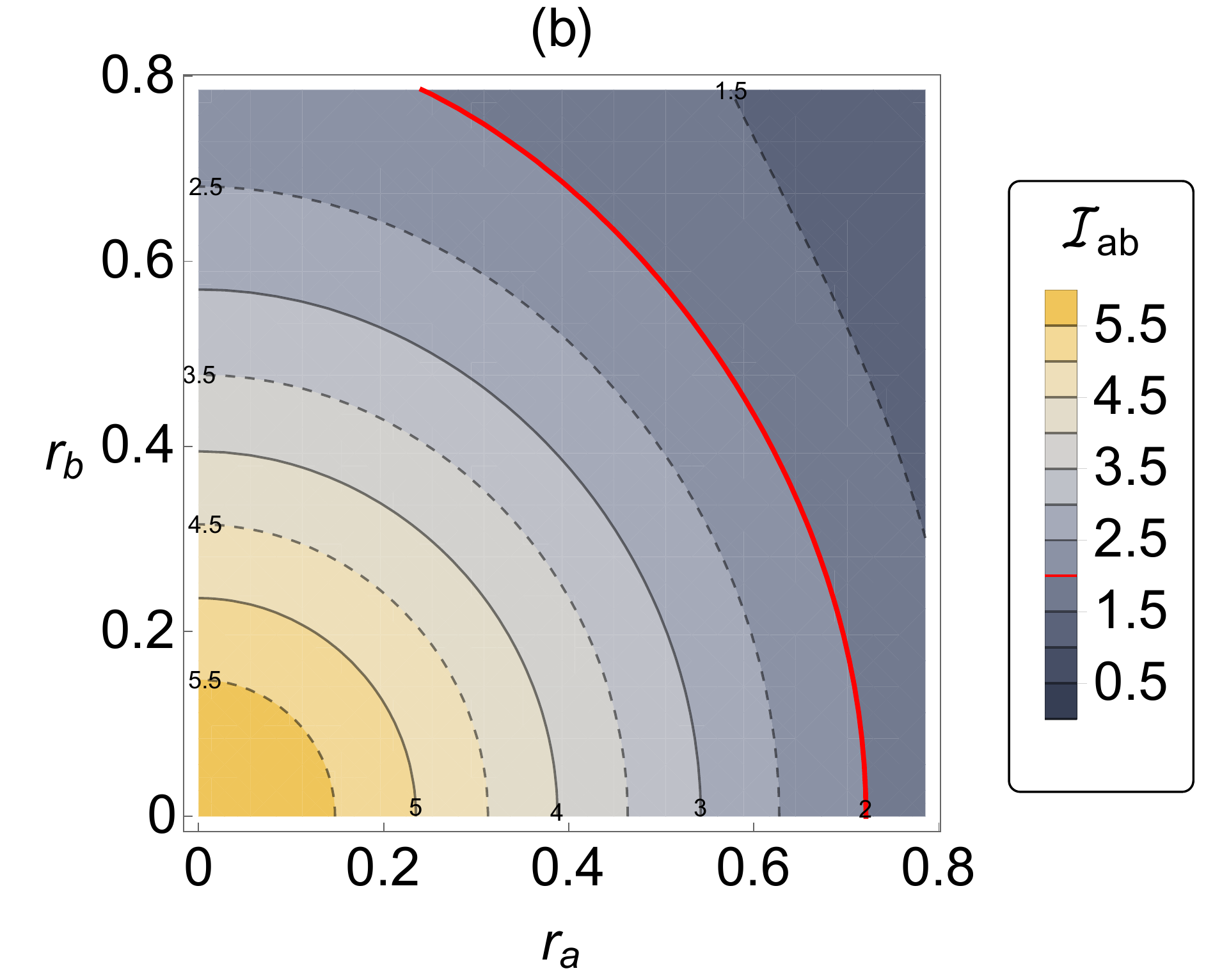}
	\includegraphics[width=0.32\linewidth, height=3.7cm]{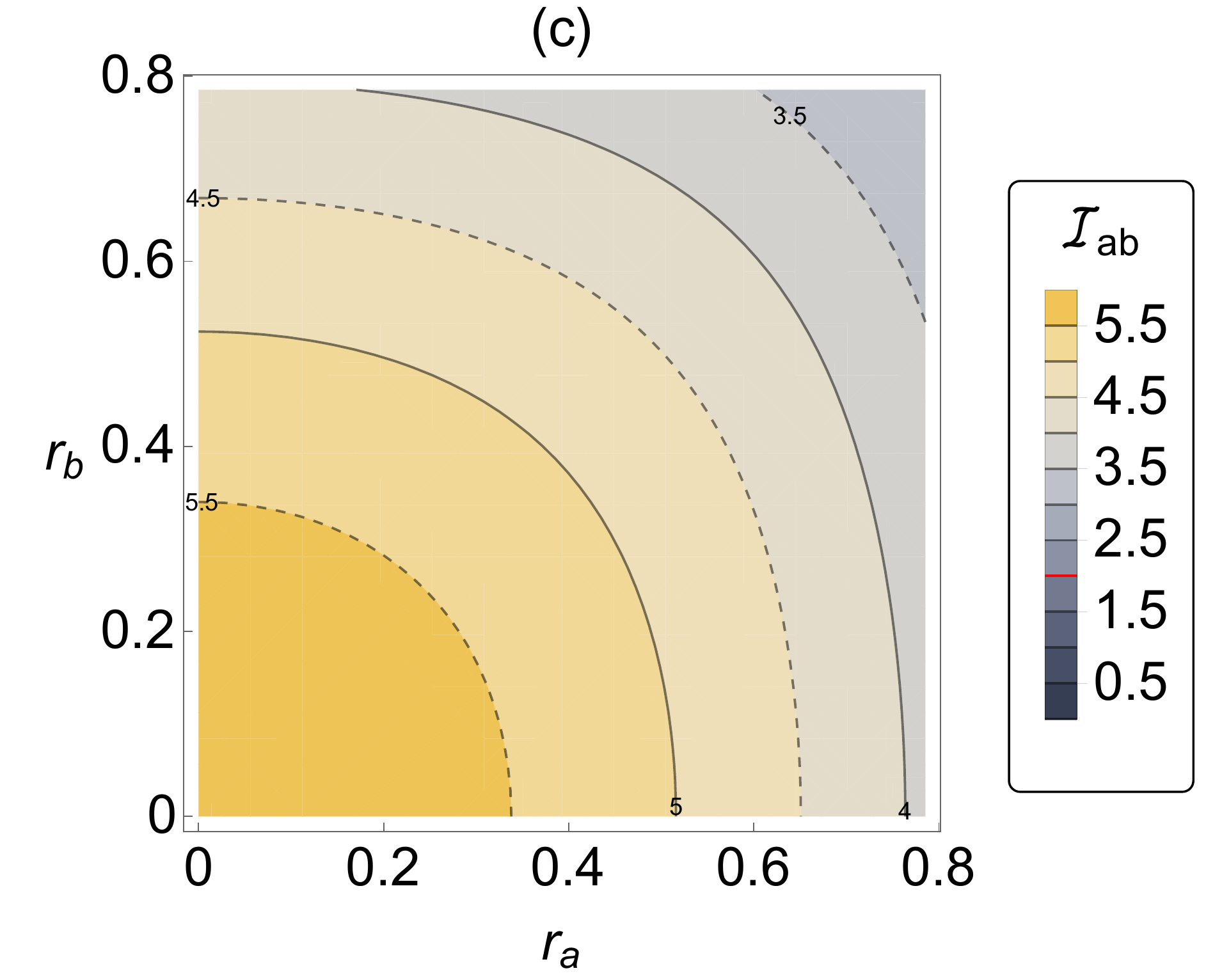}
	\includegraphics[width=0.32\linewidth, height=3.7cm]{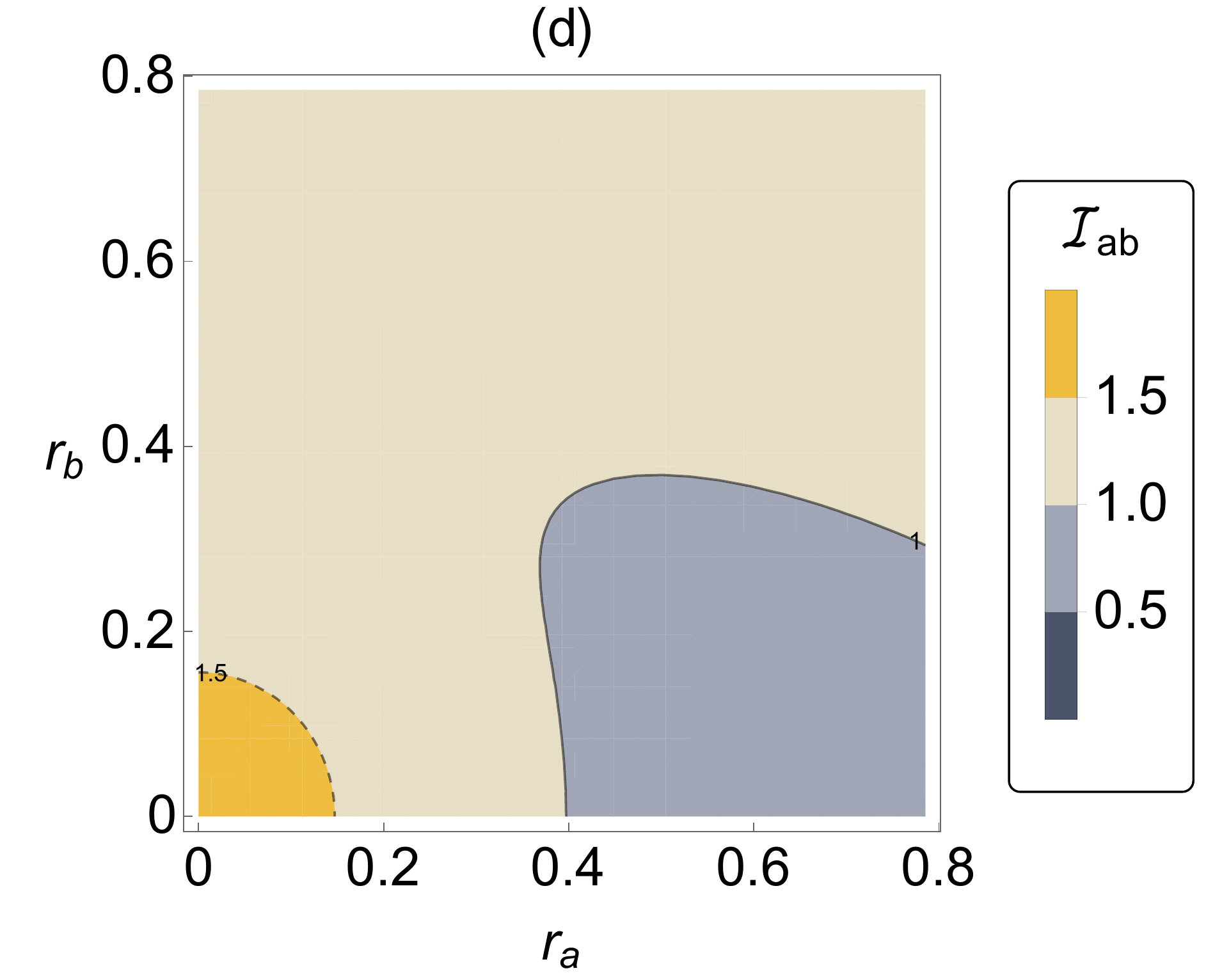}
	\includegraphics[width=0.32\linewidth, height=3.7cm]{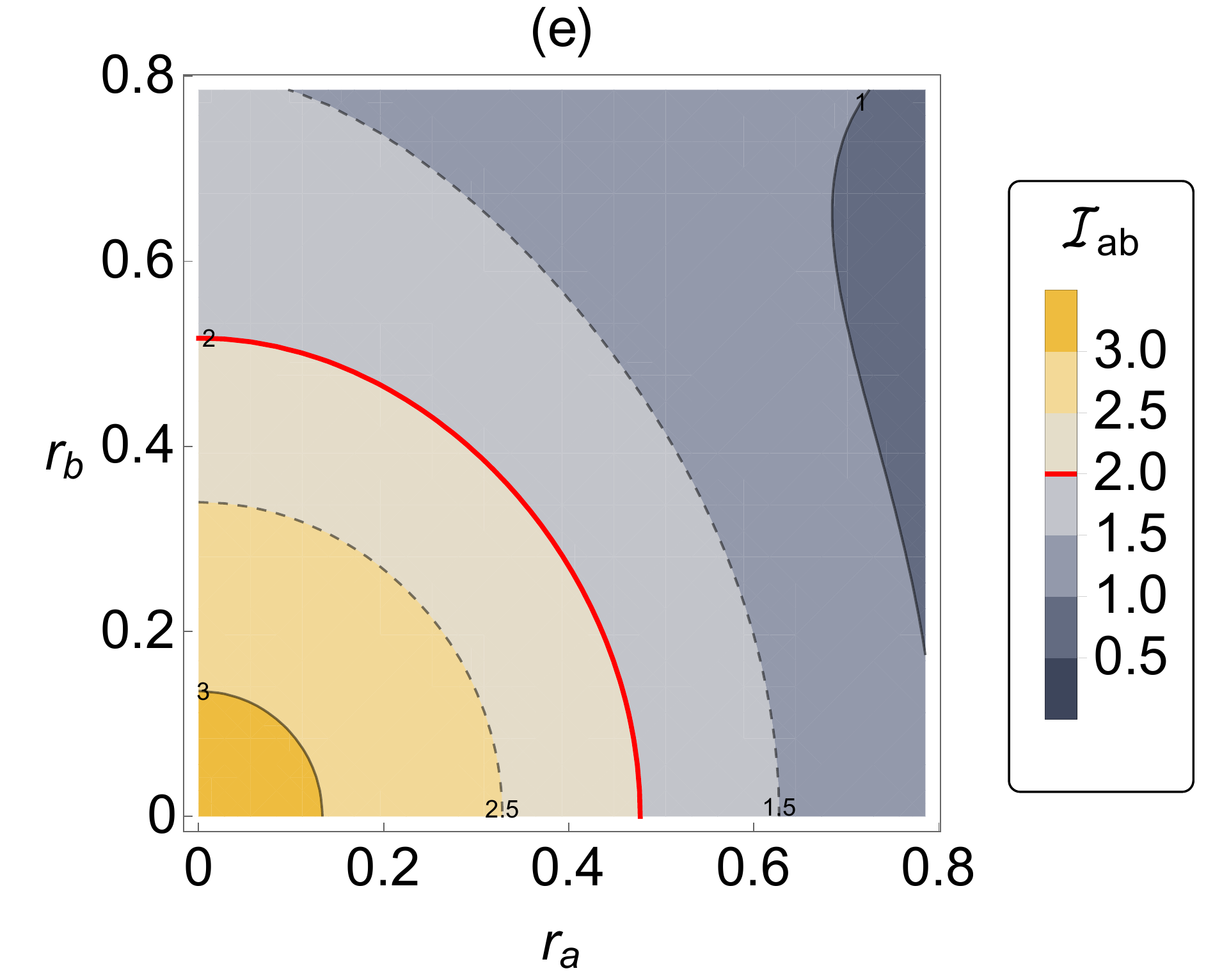}
	\includegraphics[width=0.32\linewidth, height=3.7cm]{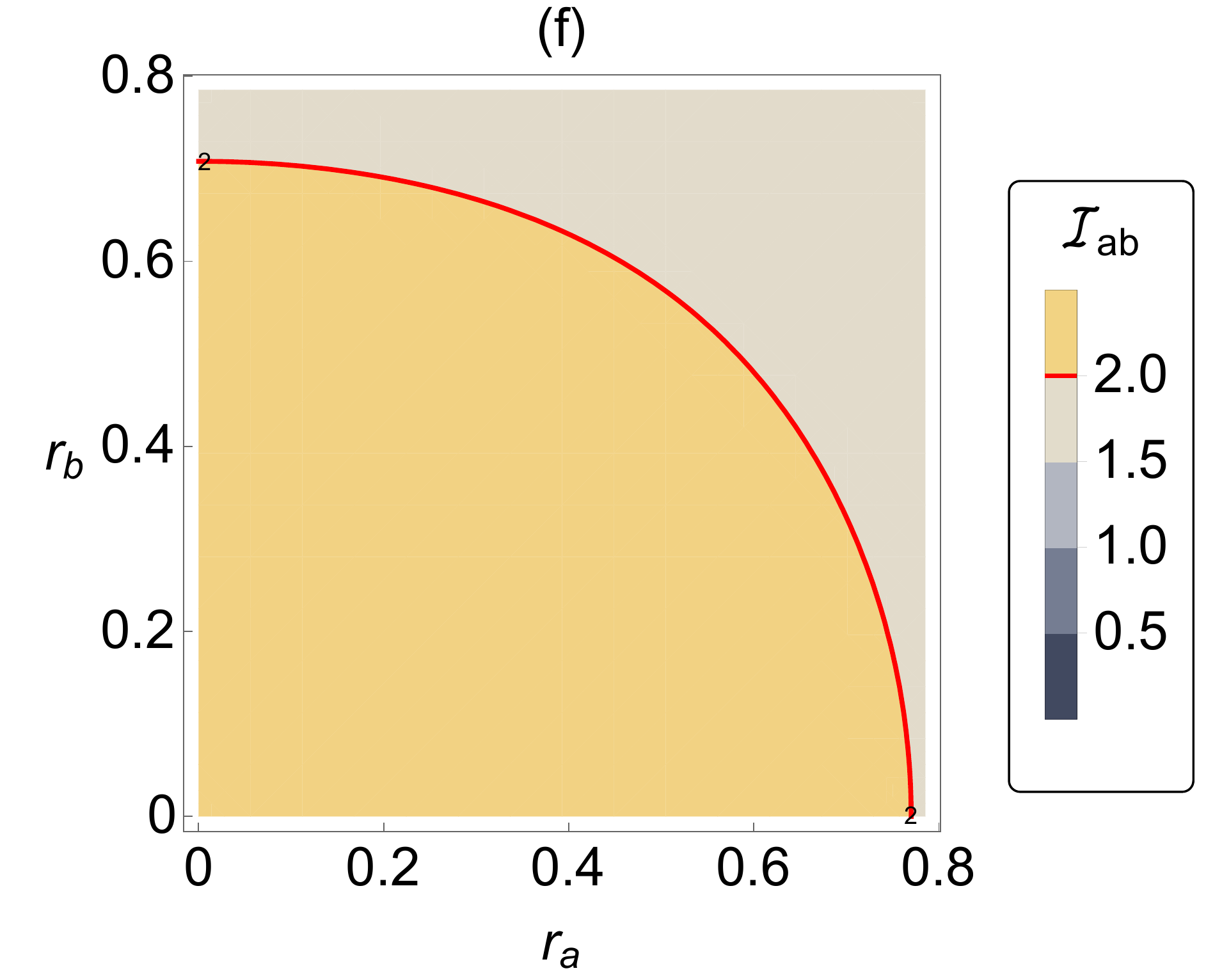}
	\caption{Steering from Alice to Bob with  $c_{11}=c_{22}=c_{33}=-1$ (a) $ \alpha=0.1 $, (b) $ \alpha=0.4 $ .(c)$ \alpha=0.7$. (d,e,f) the same as (a,b,c) respectively, but $c_{11}=c_{22}=c_{33}=-0.8$.}
	\label{Fig1-Max}
\end{figure}
In Fig.(\ref{Fig1-Max}), the behavior of steerability is displayed, where we consider that the two qubit systems are initially prepared in the  singlet state, namely we set $c_{11}=c_{22}=c_{33}=-1$.  Due to the symmetry, the behavior of $\mathcal{I}_{ab}$ is the same as $\mathcal{I}_{ba}$,  thus we consider only   $\mathcal{I}_{ab}$. In this discussion, it is assumed  that, both  qubits  are accelerated.  Fig.(\ref{Fig1-Max}.a), shows the behavior of the steering inequality $\mathcal{I}_{ab}$ before  applying the filtering process.  The general behaviour shows that, the steerability is violated as the acceleration increases.  However, it is well known that, the decoherence  of the singlet state  due to the acceleration is bounded, where the minimum  entanglement is $\sim 0.3$ \cite{45}. Therefore the possibility  that Alice steers Bob's state is violated at $r_a(r_b)>0.4$.
The effect of the filtering process  on the steerability is displayed in Fig.(\ref{Fig1-Max}b), and Fig.(\ref{Fig1-Max}c), where we set $\alpha=0.4$ and $0.7$, respectively. The behavior  of $\mathcal{I}_{ab}$ shows that, the steerability could be implemented  at any value of the acceleration.

The effect of the initial state settings is displayed in Figs. (\ref{Fig1-Max}d-\ref{Fig1-Max}f), where  it is  assumed that the users share Werner state, such that $c_{11}=c_{22}=c_{33}=-0.8$.  The behavior of the steerability  is similar to that displayed for the singlet state, but it is  violated at smaller values of the accelerations.   As it is displayed from Fig.(\ref{Fig1-Max}d), the inequality of steering is violated at $r_a=r_b< 0.2$. By applying the filtering process, the steerability could be implemented at large values of the accelerations as it  is  displayed  from Figs.(\ref{Fig1-Max}e) and (\ref{Fig1-Max}f), where the filtering strength is $0.4$ and $0.7$, respectively.

\begin{figure}[h!]
	\centering
	\includegraphics[width=0.32\linewidth, height=3.7cm]{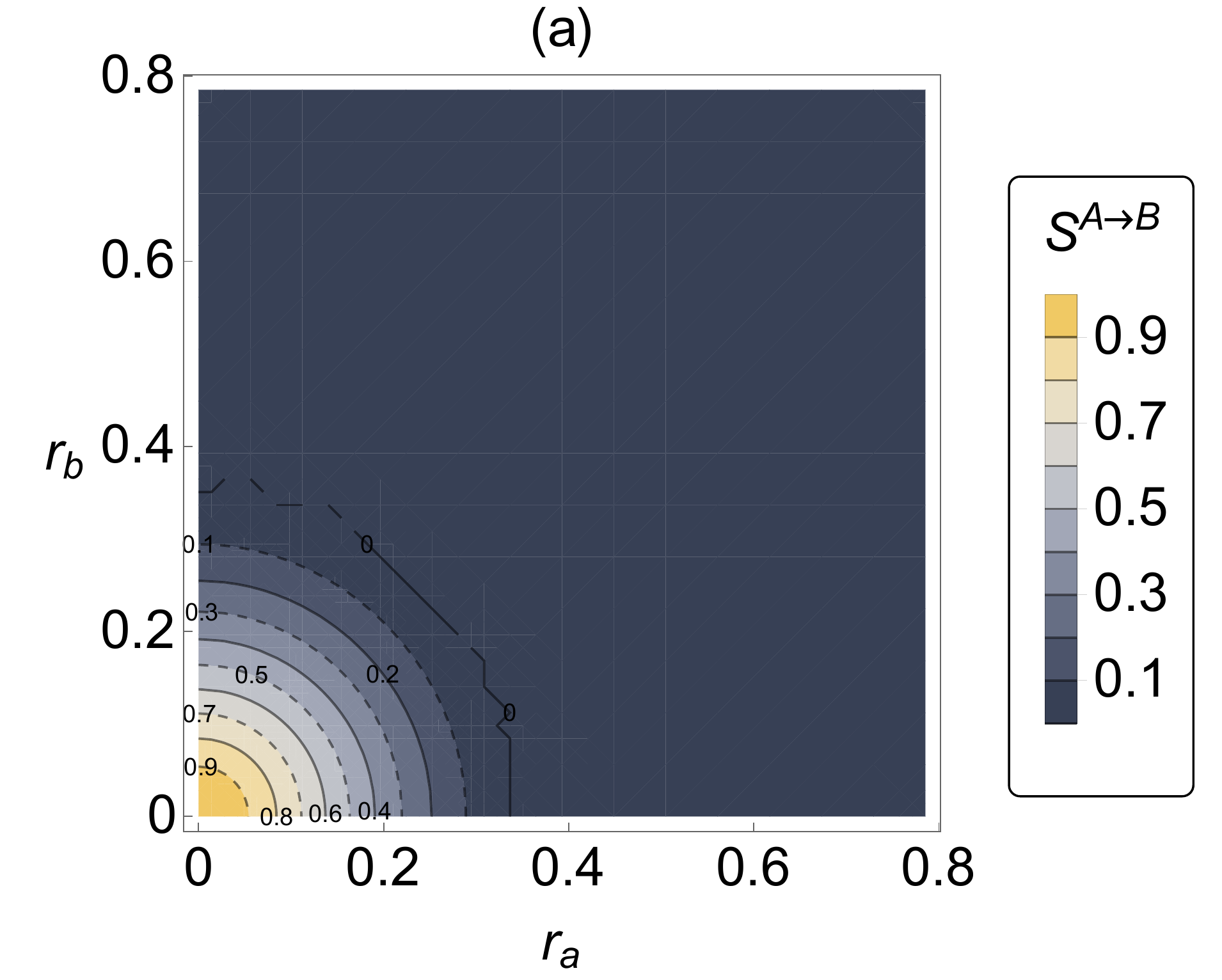}
	\includegraphics[width=0.32\linewidth, height=3.7cm]{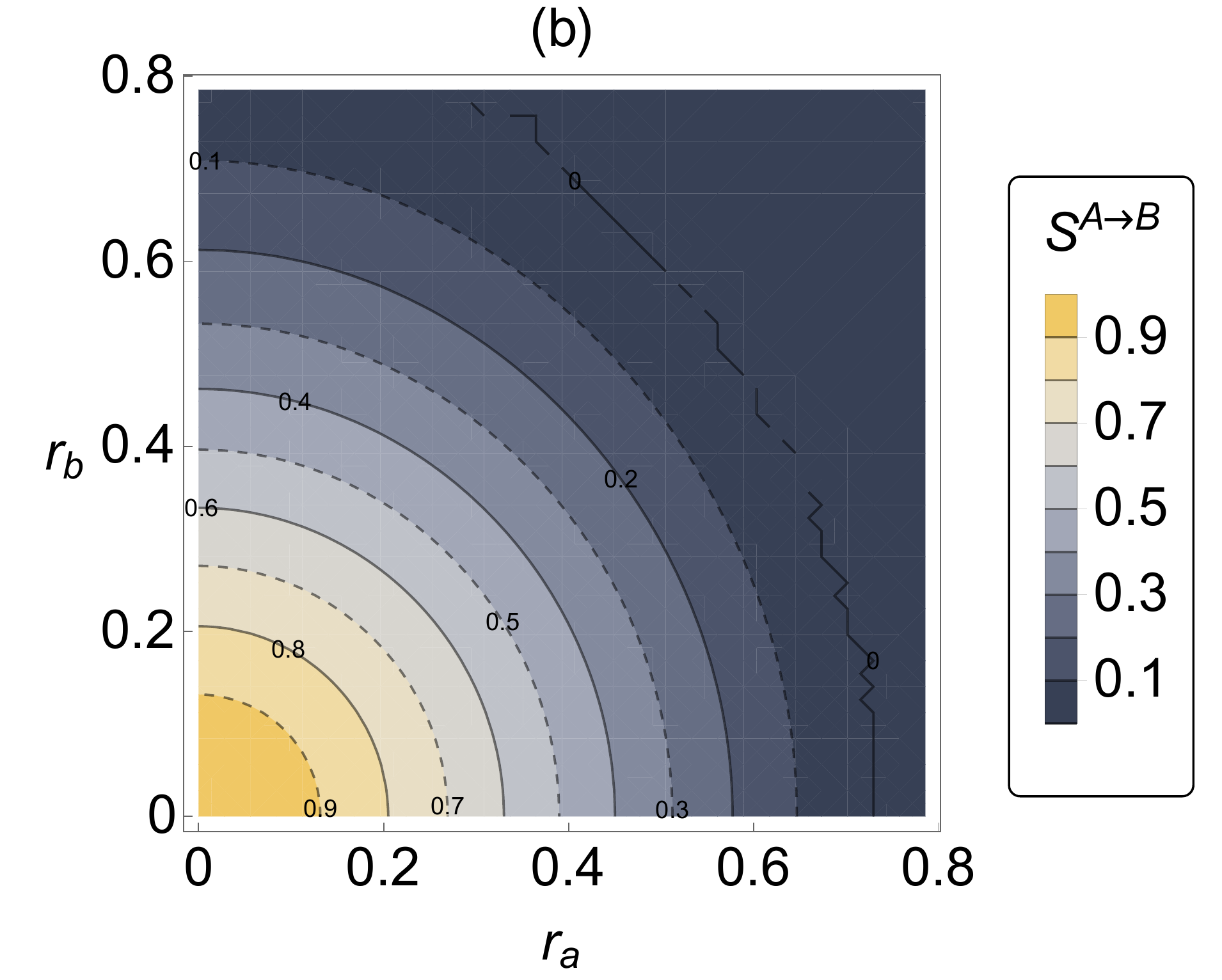}
	\includegraphics[width=0.32\linewidth, height=3.7cm]{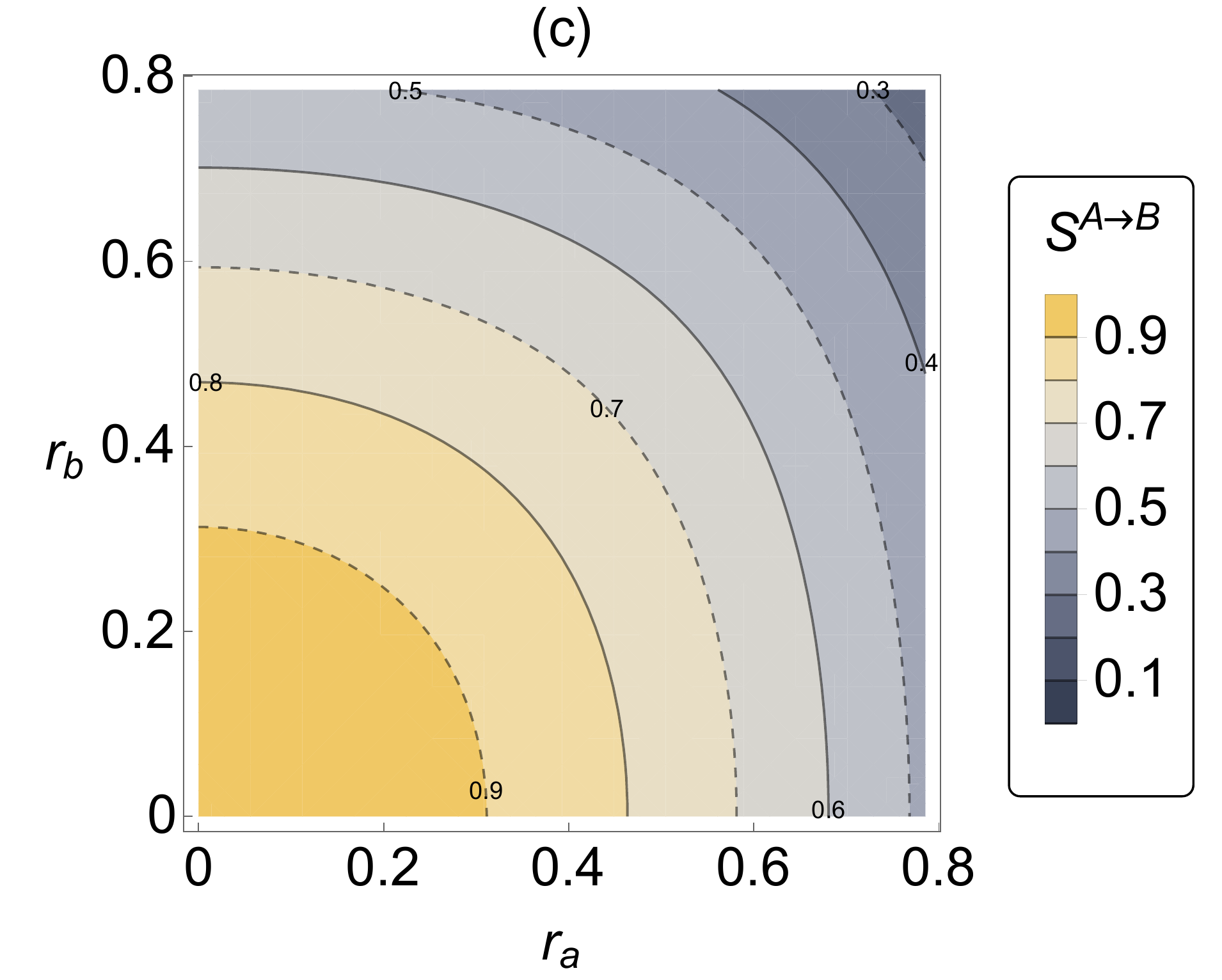}
	\includegraphics[width=0.32\linewidth, height=3.7cm]{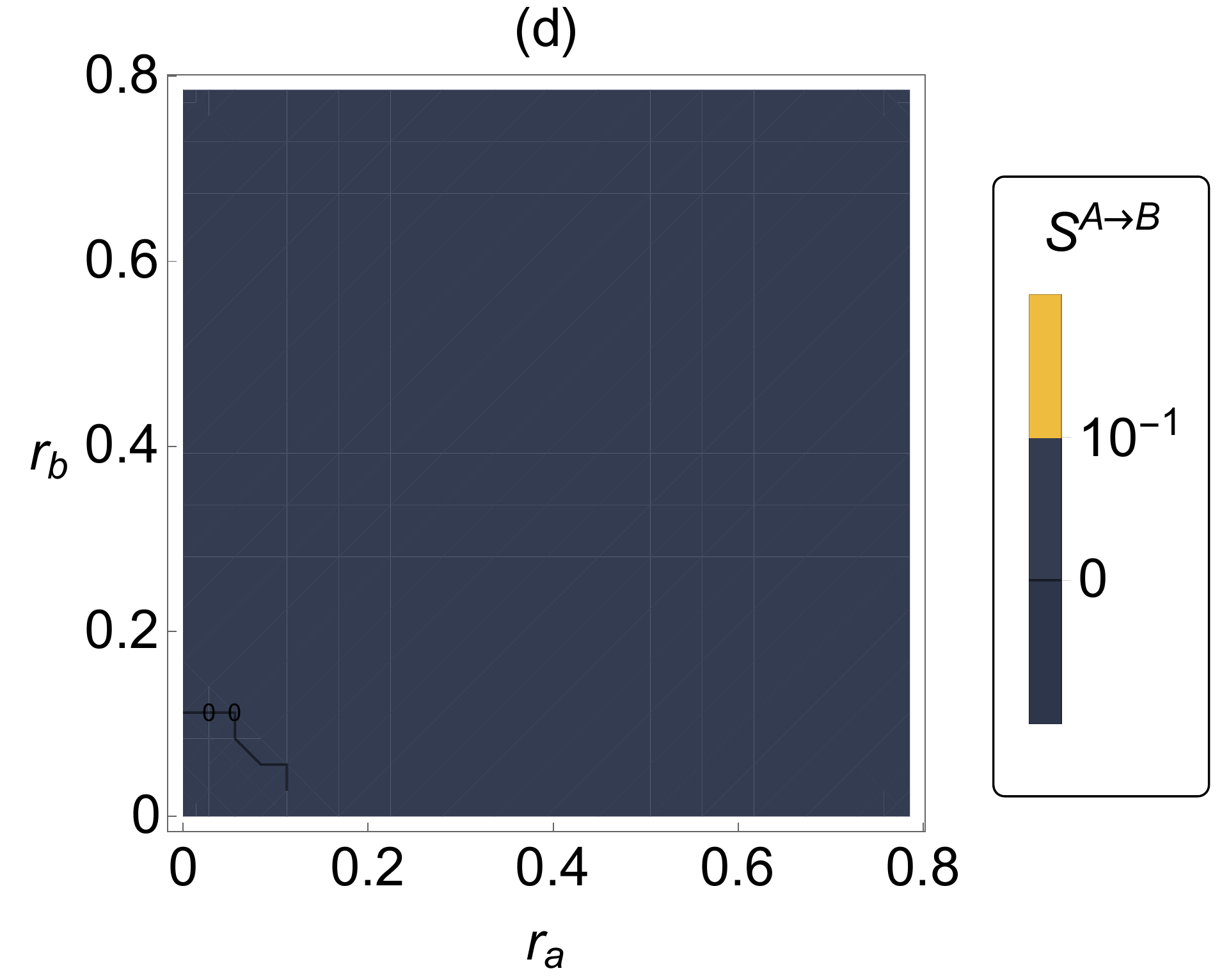}
	\includegraphics[width=0.32\linewidth, height=3.7cm]{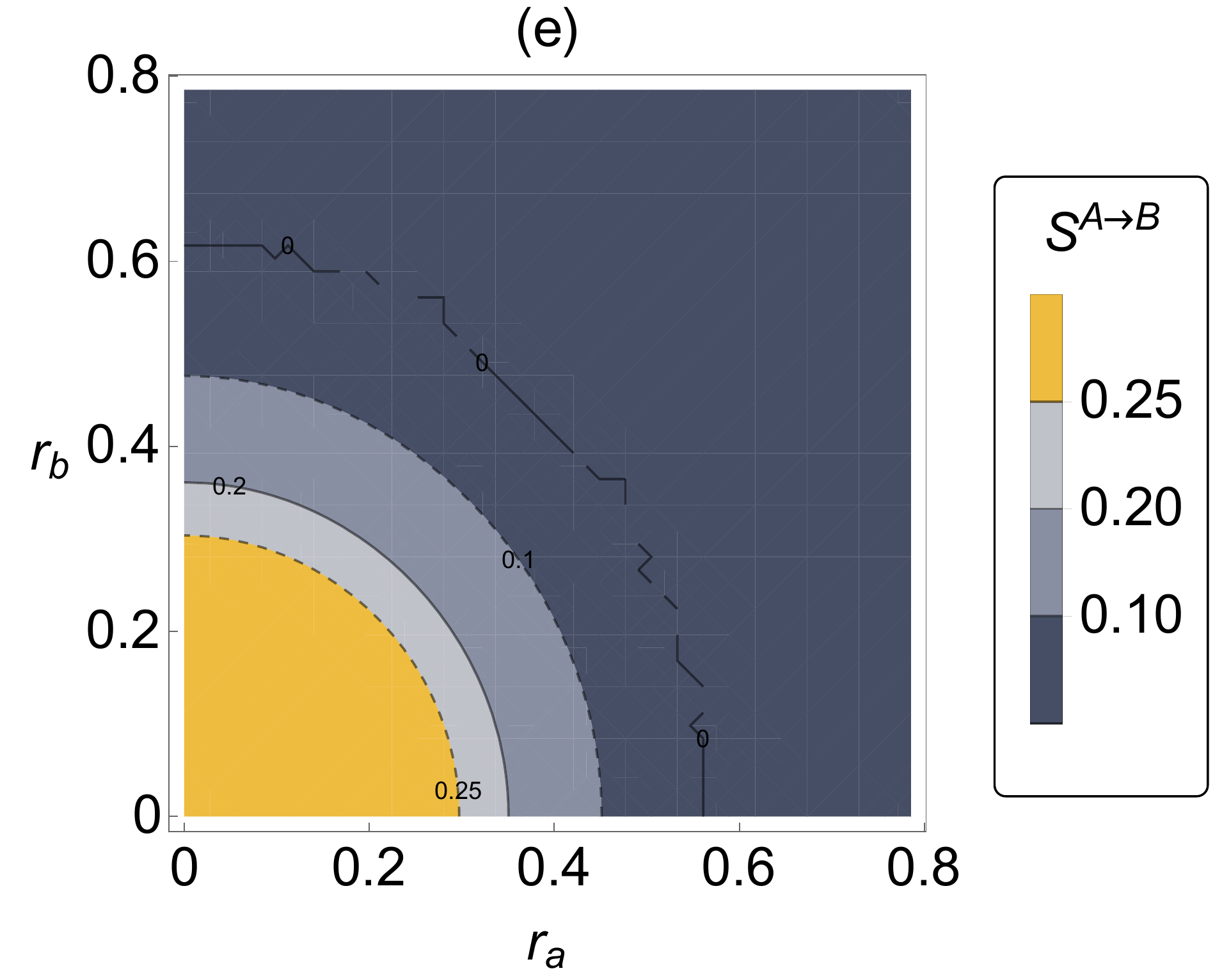}
    \includegraphics[width=0.32\linewidth, height=3.7cm]{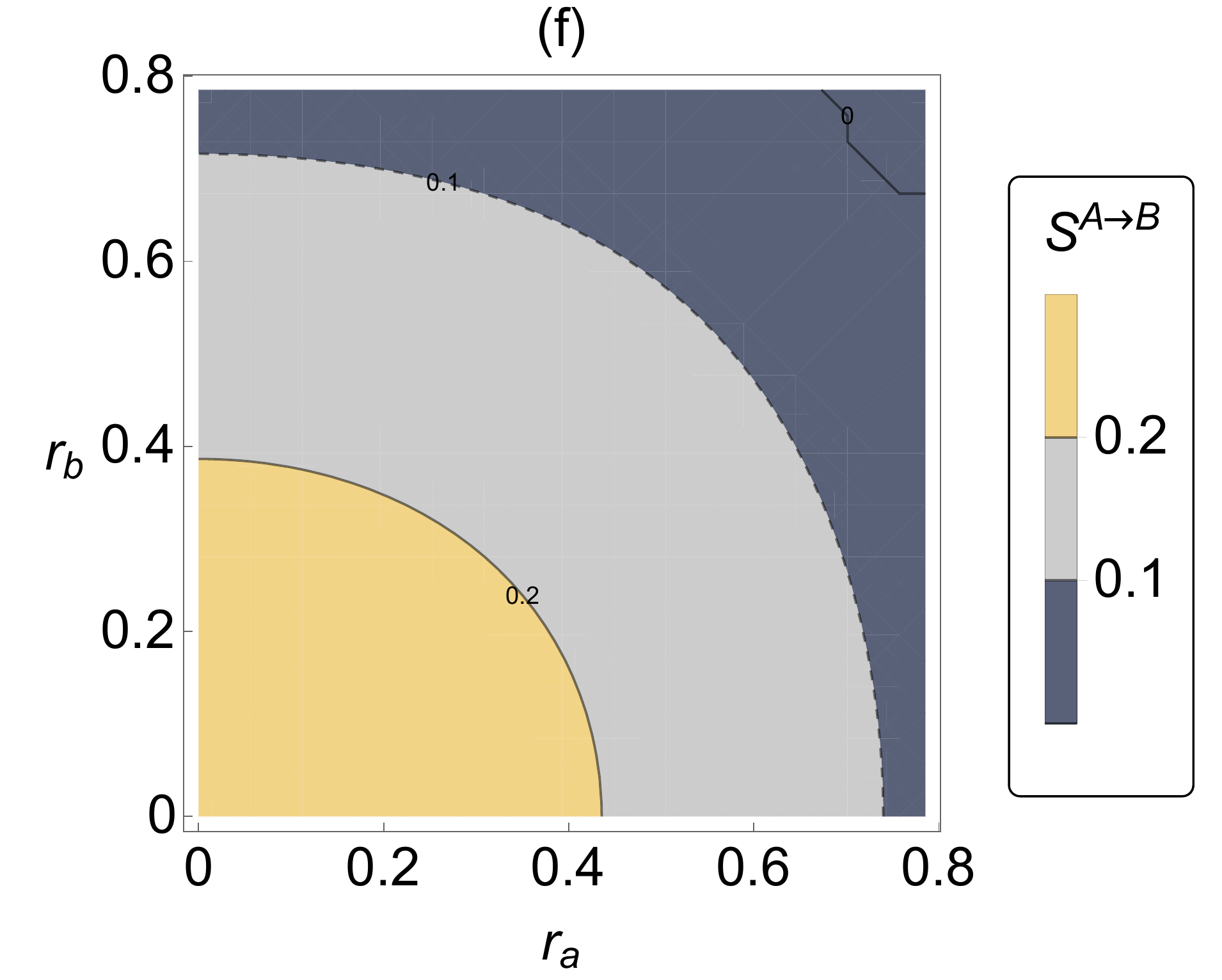}
	\caption{ The same as Fig.(\ref{Fig1-Max}),  but for the degree of steerability $\mathcal{S}^{A\longrightarrow B}$. }
	\label{Fig2SM}
\end{figure}

The degree of steerability  $\mathcal{S}^{A\longrightarrow B}$ that Alice can steer Bob's qubit is displayed in Fig.(\ref{Fig2SM}), where it is  assumed that,  the steerer and the steered partner share a singlet state. The general behavior of  $\mathcal{S}^{A\longrightarrow B}$ shows that, the degree of steerability decreases as the acceleration increases. In the absence of the filtering process, the  degree of the steerability decreases fast, while when the filtering process is applied, the steerability decreases slowly as the acceleration of both qubits increases.  As it is displayed from Fig.(\ref{Fig2SM}b) and Fig.(\ref{Fig2SM}c), where we set  $\alpha=0.4$, and  $ 0.7$, respectively,  the degree of steerability that is depicted at $r_a=r_b=0.6$ is given by   $\mathcal{S}^{A\longrightarrow B}=0.2$ and  $0.7$, respectively.

In Figs.(\ref{Fig2SM}d-\ref{Fig2SM}f) we quantify the degree of steerability $\mathcal{S}^{A\longrightarrow B}$, where the partners share a Werner state. It is clear that, its upper bounds are smaller than those displayed for the singlet state. The filtering process has the same effect, namely it increases as one increases the filtering strength $\alpha$.

\subsection{Generic Pure state:}
For this state, the coherence (Bloch) vectors  of both qubits are equals, namely, $|\vec{r}|=p=|\vec{s}|$. The non-zero elements in the correlation matrix are defined by $ c_{11}=-1$, and $ c_{22}=-q=c_{33}$. In an explicit form, the state (\ref{state}) may be written as:
 \begin{equation}
 	\hat{\rho}_{ab}=\frac{1}{4}\bigg(I^{(a)}_2\otimes I^{(b)}_2-\sigma_x\otimes \tau_x+p(\sigma_x\otimes I^{(b)}_2-I^{(a)}_2\otimes \tau_x)-q(\sigma_y\otimes \tau_y+\sigma_z\otimes \tau_z)\bigg),
 \end{equation}
 where, $p=\sqrt{1-q^2}$. After the acceleration process, the generic accelerated  pure state is given by,
 \begin{equation}
 		\begin{split}
 	\hat{\rho}_{AB}^{acc}= &\mathcal{C}^2_a \mathcal{C}^2_b\mathcal{B}_{11}|00\rangle \langle 00|+ \mathcal{C}^2_a \big(\mathcal{B}_{22}+ S^2_b \mathcal{B}_{11}\big) |01\rangle \langle 01|+
 	\mathcal{C}^2_b \big(\mathcal{B}_{22}+ S^2_a \mathcal{B}_{11}\big)|10\rangle \langle 10|+\\& (S^2_a \big(\mathcal{B}_{22}+ S^2_b \mathcal{B}_{11}\big)+ S^2_b \mathcal{B}_{22}+\mathcal{B}_{11})|11\rangle \langle 11|+ \mathcal{C}_a \mathcal{C}_b \big\{\mathcal{B}_{14}|00\rangle \langle 11|+\mathcal{B}_{23}|01\rangle \langle 10|+\\&\mathcal{B}_{12}\mathcal{C}_a\big(-|00\rangle \langle 01|+(1+S_b^2)|01\rangle \langle 11|\big)+\mathcal{B}_{12}\mathcal{C}_b\big(|00\rangle \langle 10|-(1+S_a^2)|10\rangle \langle 11|\big)
 	+h.c.\big\},
 	\end{split}
 \end{equation}
 with,
 \begin{equation}
 	\mathcal{B}_{11}=\frac{1-q}{4}=-\mathcal{B}_{14}, \quad 	\mathcal{B}_{22}=\frac{1+q}{4}=-\mathcal{B}_{23}, \quad	 \mathcal{B}_{12}=\frac{p}{4}.
 \end{equation}

\begin{figure}[!h]
	\centering
	\includegraphics[width=0.28\linewidth, height=4cm]{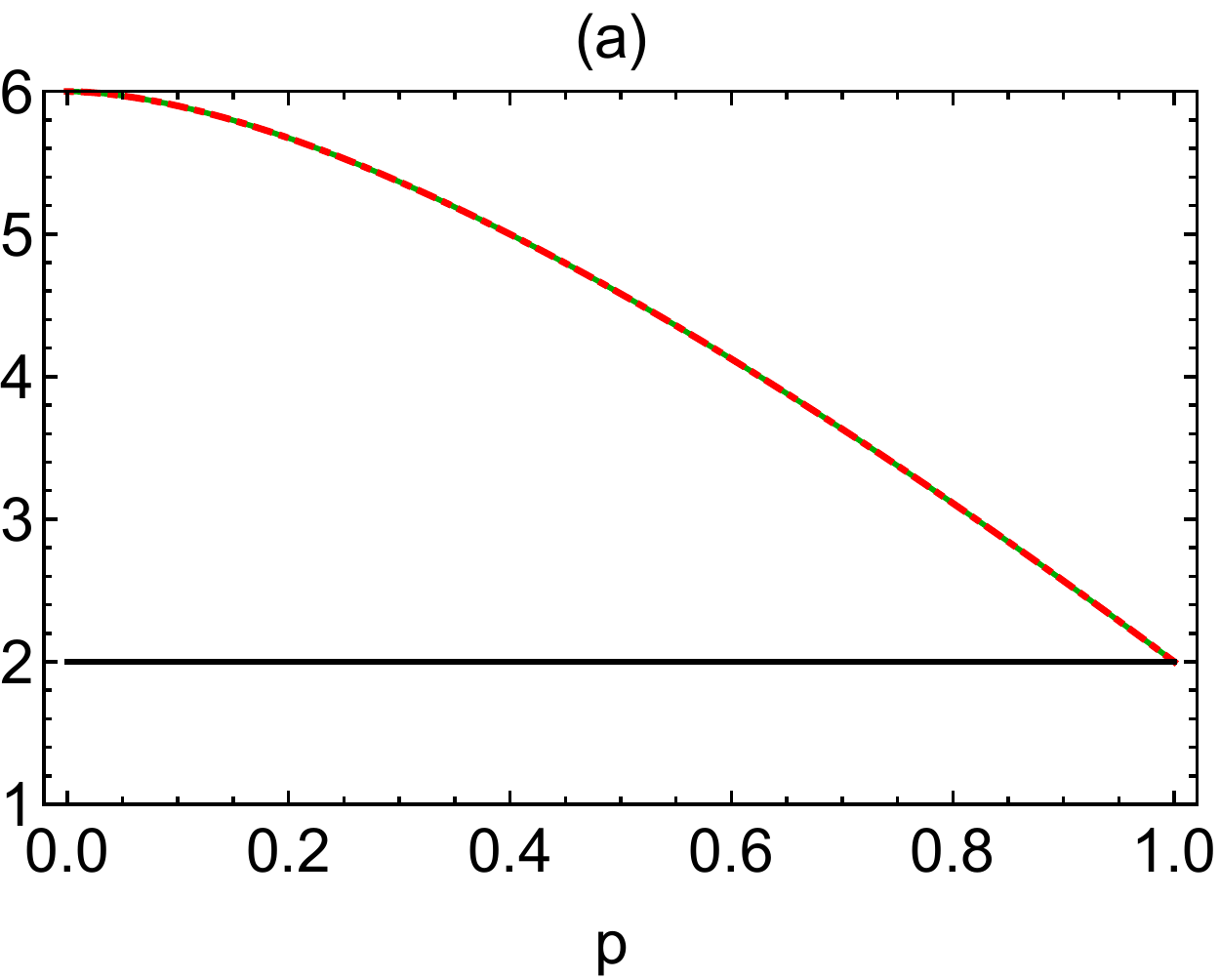}~~~\quad
	\includegraphics[width=0.28\linewidth, height=4cm]{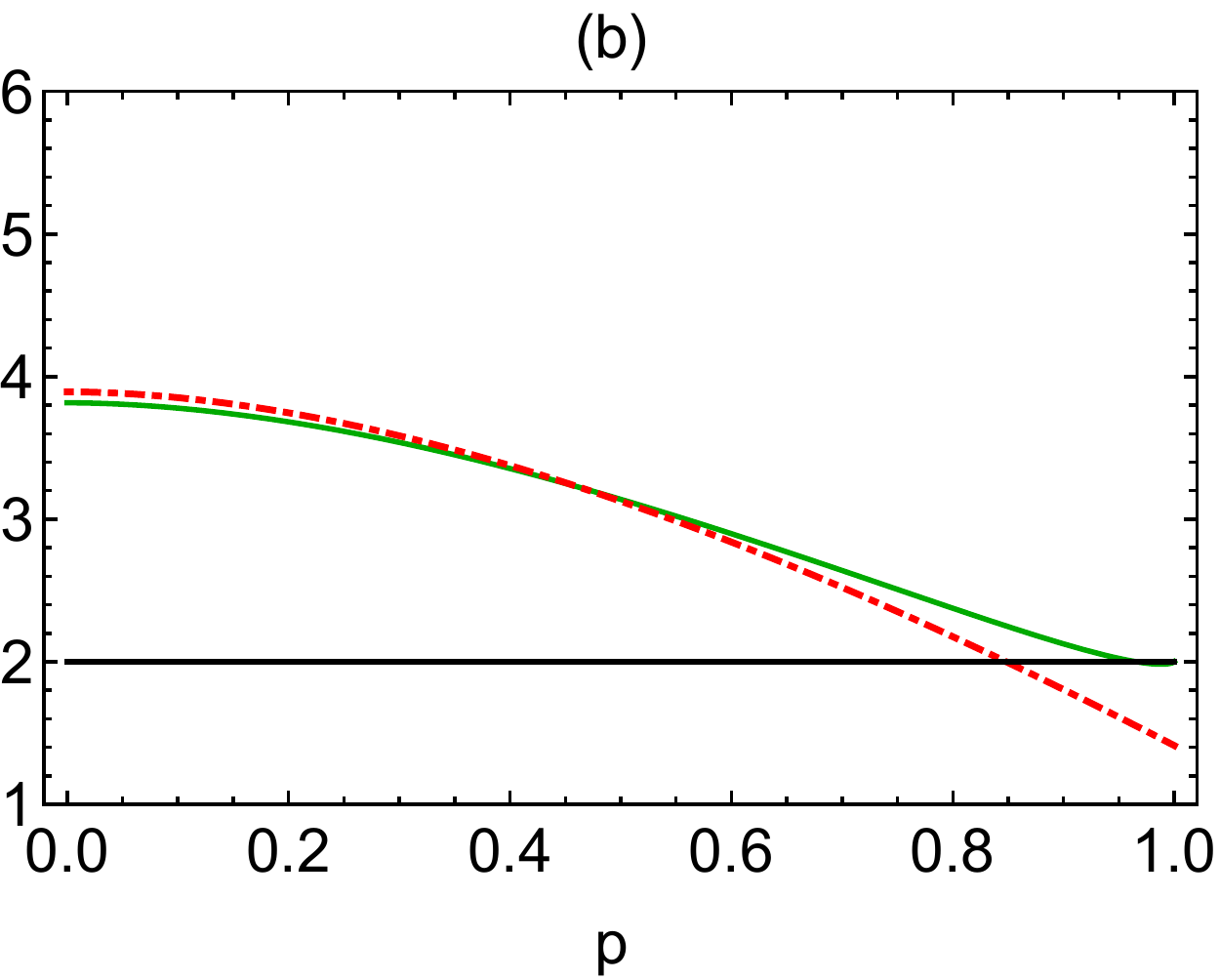}~\quad
	\includegraphics[width=0.28\linewidth, height=4cm]{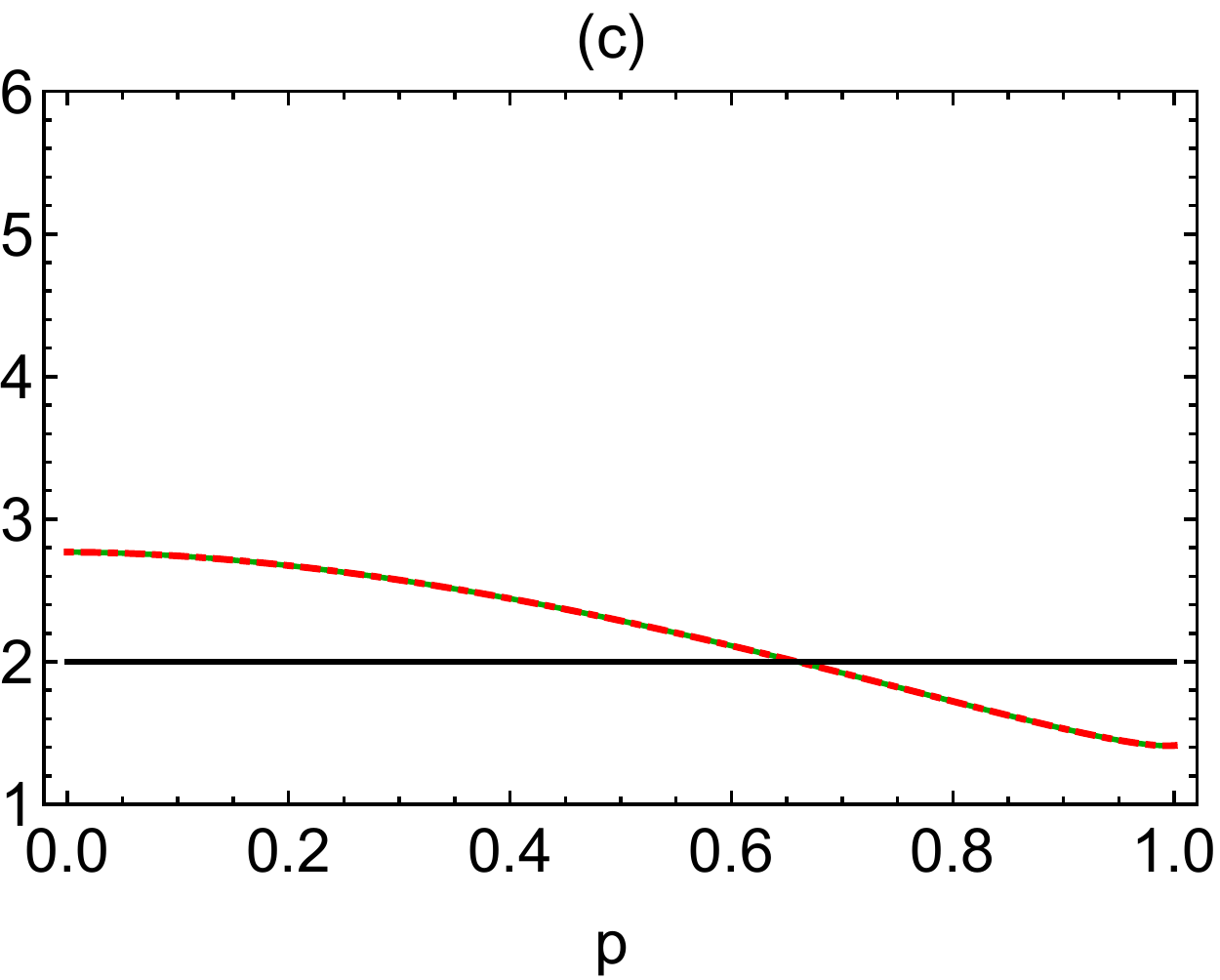}
	\caption{ quantum steering $\mathcal{I}_{ab}$ (green-solid curve), and $\mathcal{I}_{ba}$(red-dot dash curve), (a)$r_a=0=r_b$, (b)$r_a=0.5, \ r_b=0$, and (c) $r_a=0.5, \ r_b=0.5$.}
\label{Fig-I}
\end{figure}
 The inequalities of  steerability  from Alice to Bob and vise versa, are displayed in Fig.(\ref{Fig-I}),  where they decrease gradually as $p$ increases. Fig.({\ref{Fig-I}a) shows the bidirectional steering's inequalities when both the steerer and the steered  are in the  inertial frame, namely $r_a=r_b=0$. It is clear that, the behavior of both inequalities  coincides and decreases gradually as the parameter $p$ increases. The steerability   from both directions vanishes completely at $p=1$, where the initial shared state turns into a separable state.  As it is displayed from Fig.(\ref{Fig-I}.b), the steerability increases slightly  if  the steerer is accelerated and the steered particle is on the inertial frame, where $\mathcal{I}_{ab}$ satisfies the steerability criteria at large values of $p$. The consistent behavior of both inequalities is displayed in Fig.(\ref{Fig-I}c), where both users  are accelerated with the same  acceleration.

\begin{figure}[h!]
	\centering
	\includegraphics[width=0.32\linewidth, height=4cm]{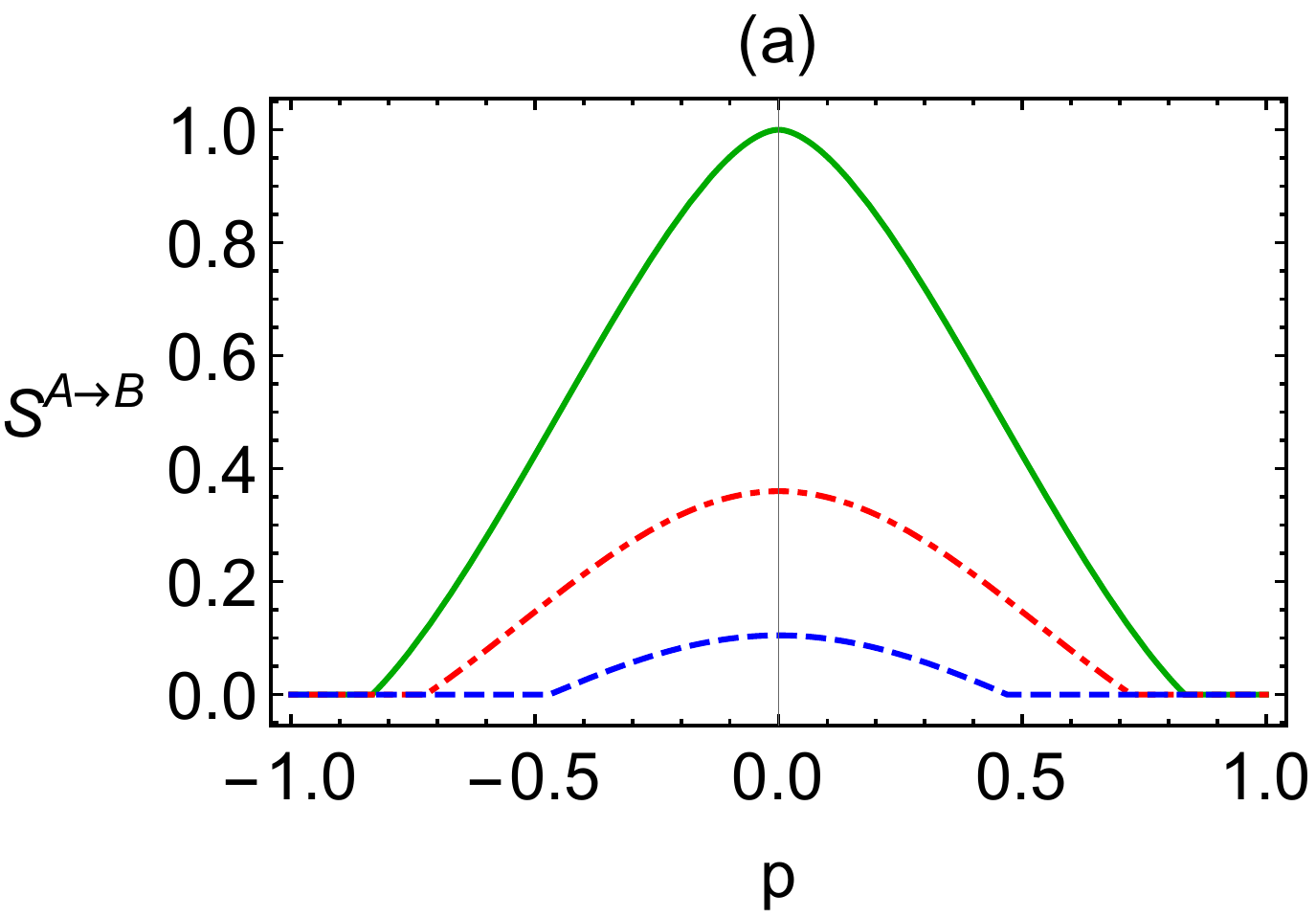}
	\includegraphics[width=0.32\linewidth, height=4cm]{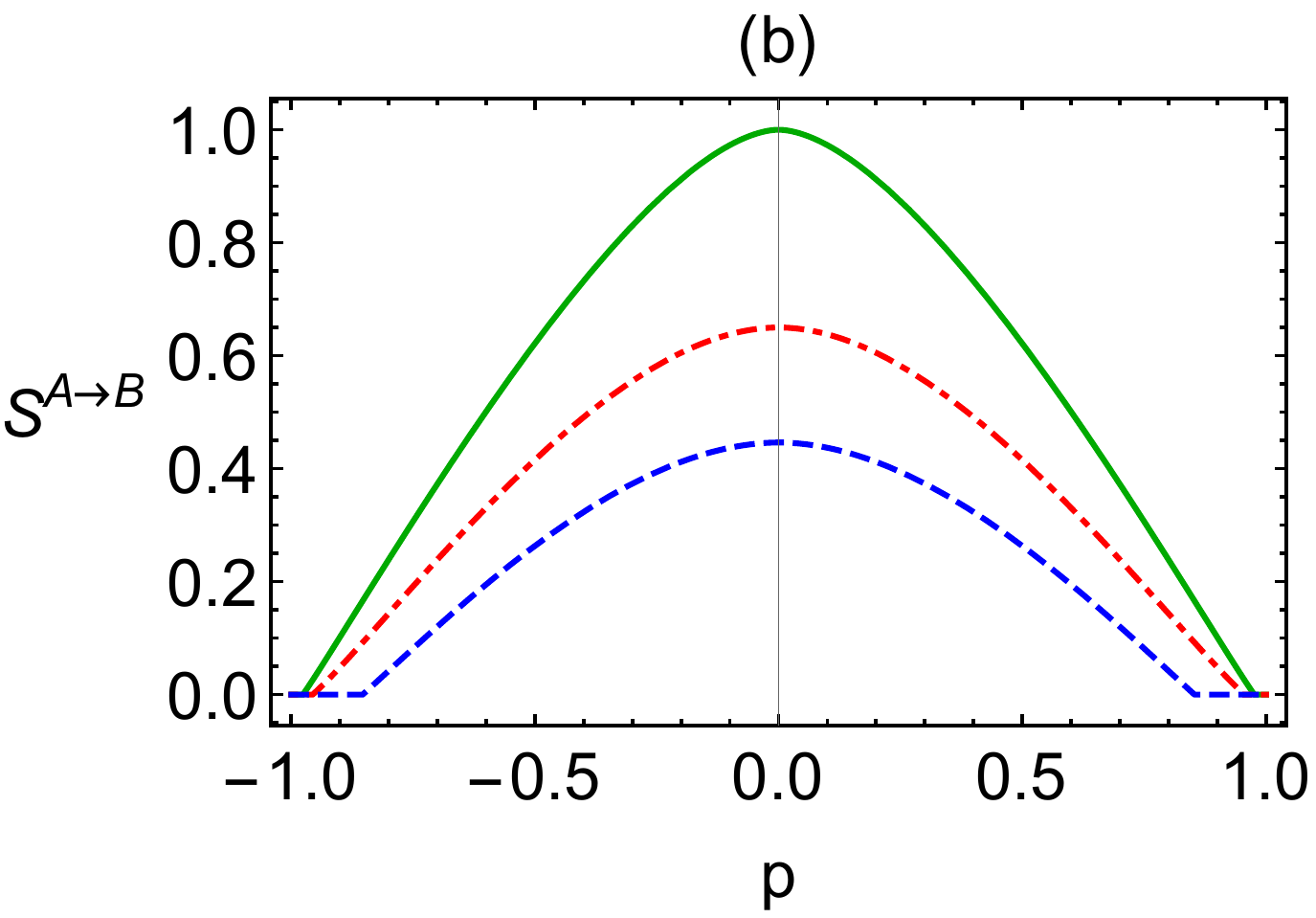}
	\includegraphics[width=0.32\linewidth, height=4cm]{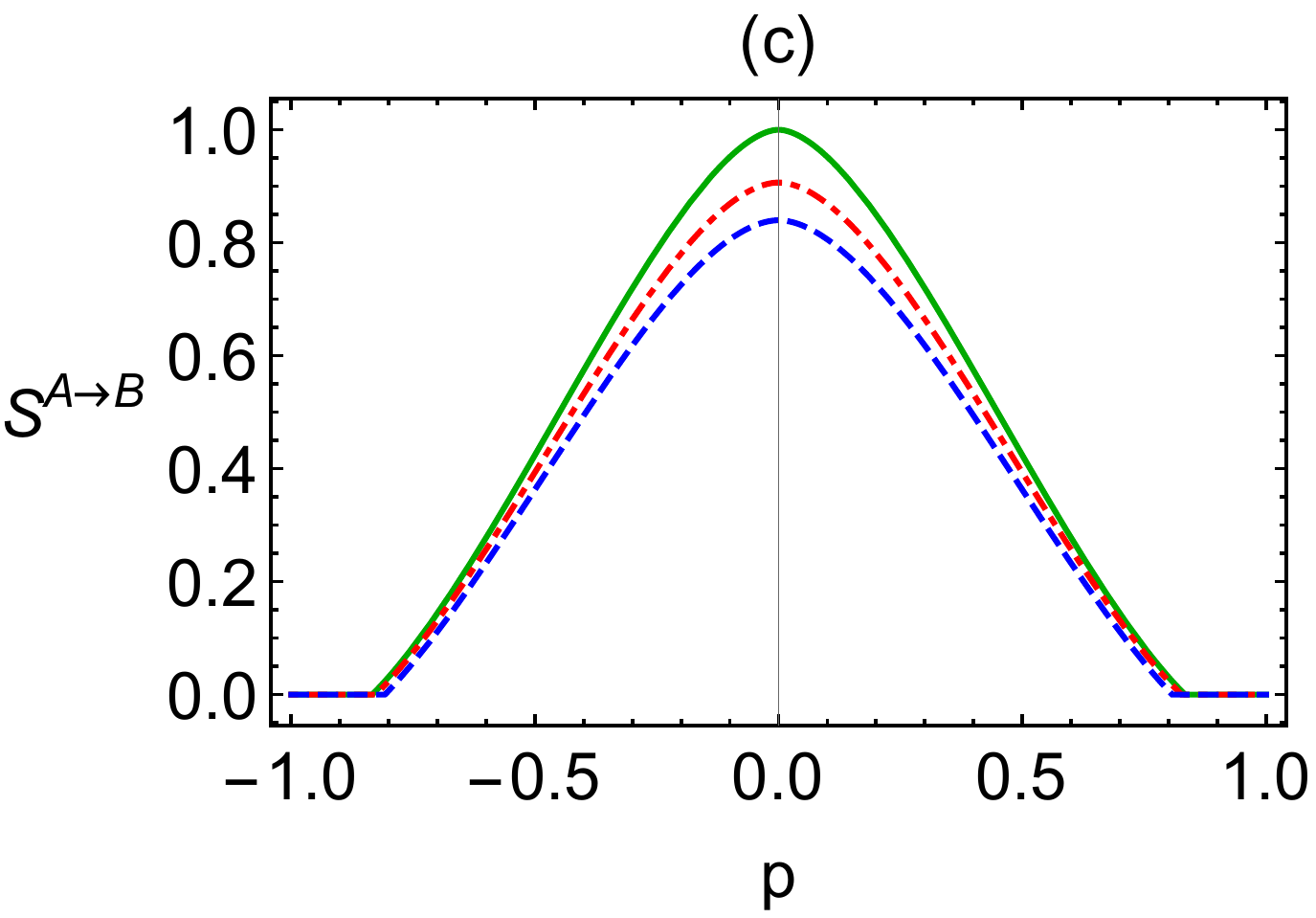}
		\caption{Steerability for generic pure state from Alice to Bob with zero acceleration (green curve), $r_a=0.3$, $r_b=0$ (red curve),  $r_a=0.3$, $r_b=0.3$ (blue curve), under weak measurement, when we set $q=\sqrt{1-p^2}$. (a)$\alpha=0.1$, (b)$\alpha=0.4$, (c)$\alpha=0.8$.}
	\label{fig:8}
\end{figure}

Fig.(\ref{fig:8}) shows the degree of steerability $\mathcal{S}^{A\longrightarrow B}$, when  the partners share  a  generic pure state. Different cases are considered, either both partners in the stationary frame, or only one partner or both partners qubits are accelerated. It is clear that, the steerability increases gradually to reach its maximum  bound at $p=0$, namely the shared state is  a maximum entangled state.  However, as $p$ increases further, the steerability decreases gradually to vanish completely. As it is displayed in Fig.(\ref{fig:8}a), the largest upper bounds are reached when the partners' qubits are in the stationary frame,  while the smallest ones are depicted when both partners are accelerated. Figs. (\ref{fig:8}b) and (\ref{fig:8}c) display the effect of the filtering process on the degree of steerability. It is clear that, as one increases the filtering strength, the upper bounds of the steerability increase.

\begin{figure}[h!]
	\centering
	\includegraphics[width=0.32\linewidth, height=4cm]{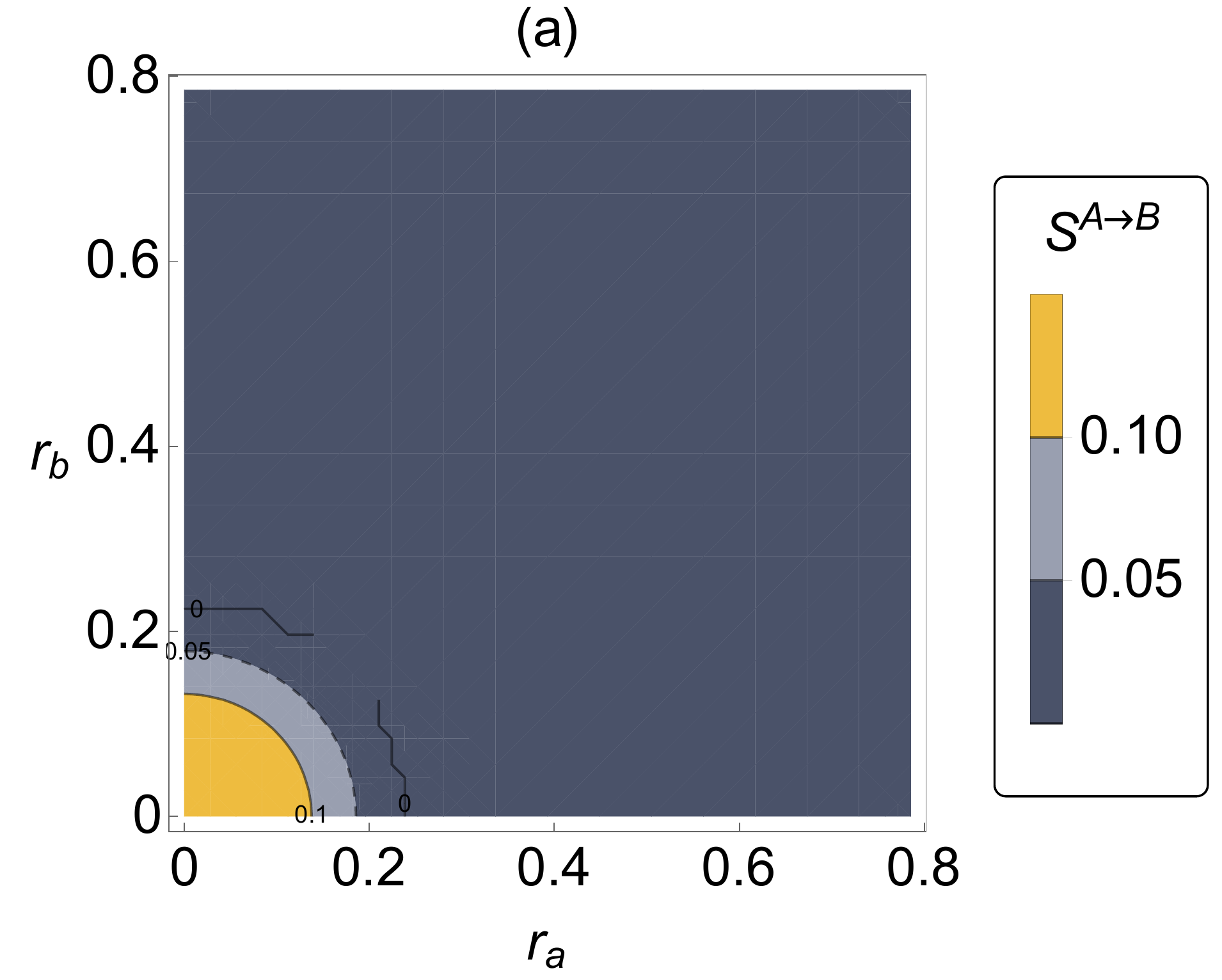}
	\includegraphics[width=0.32\linewidth, height=4cm]{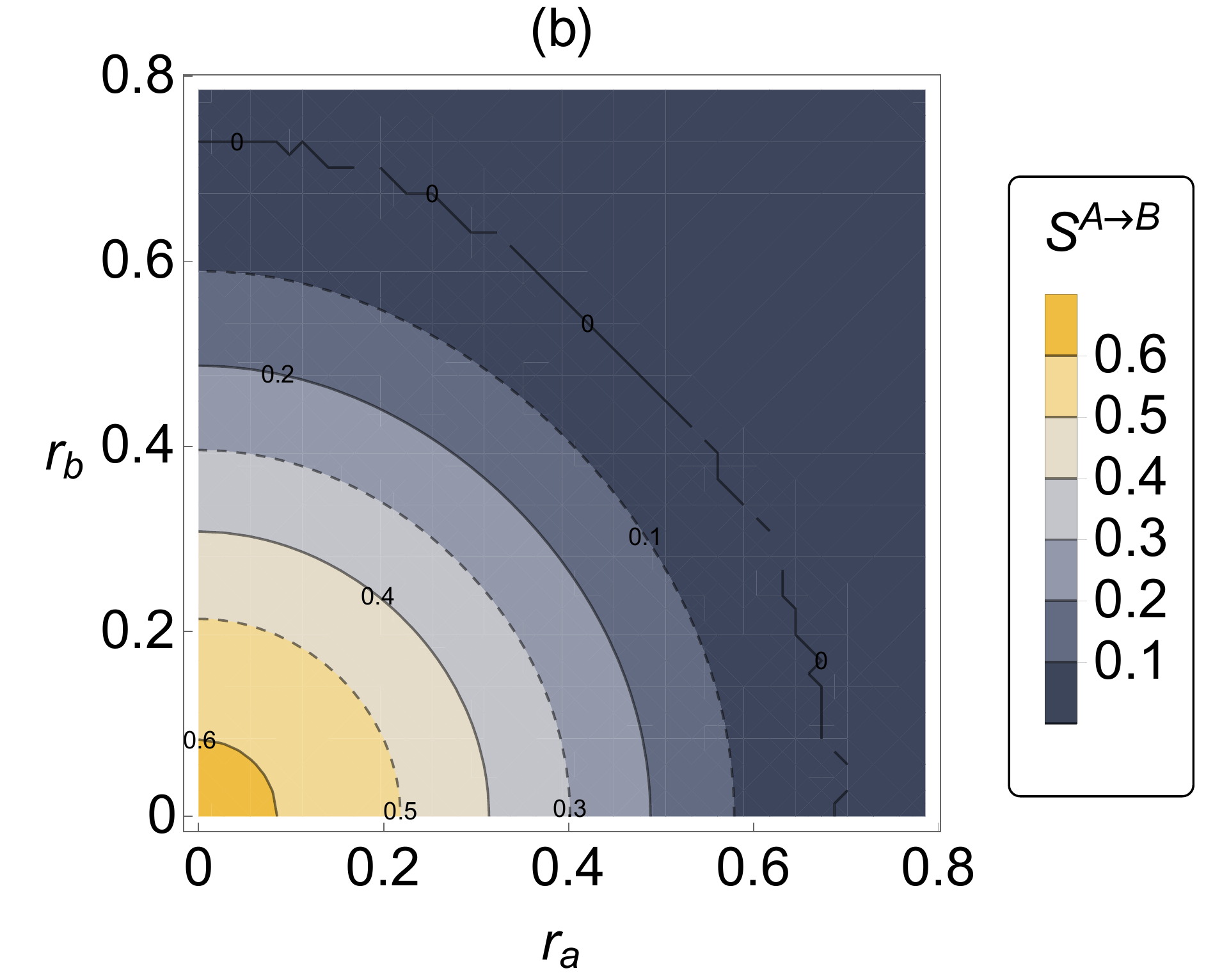}
	\includegraphics[width=0.32\linewidth, height=4cm]{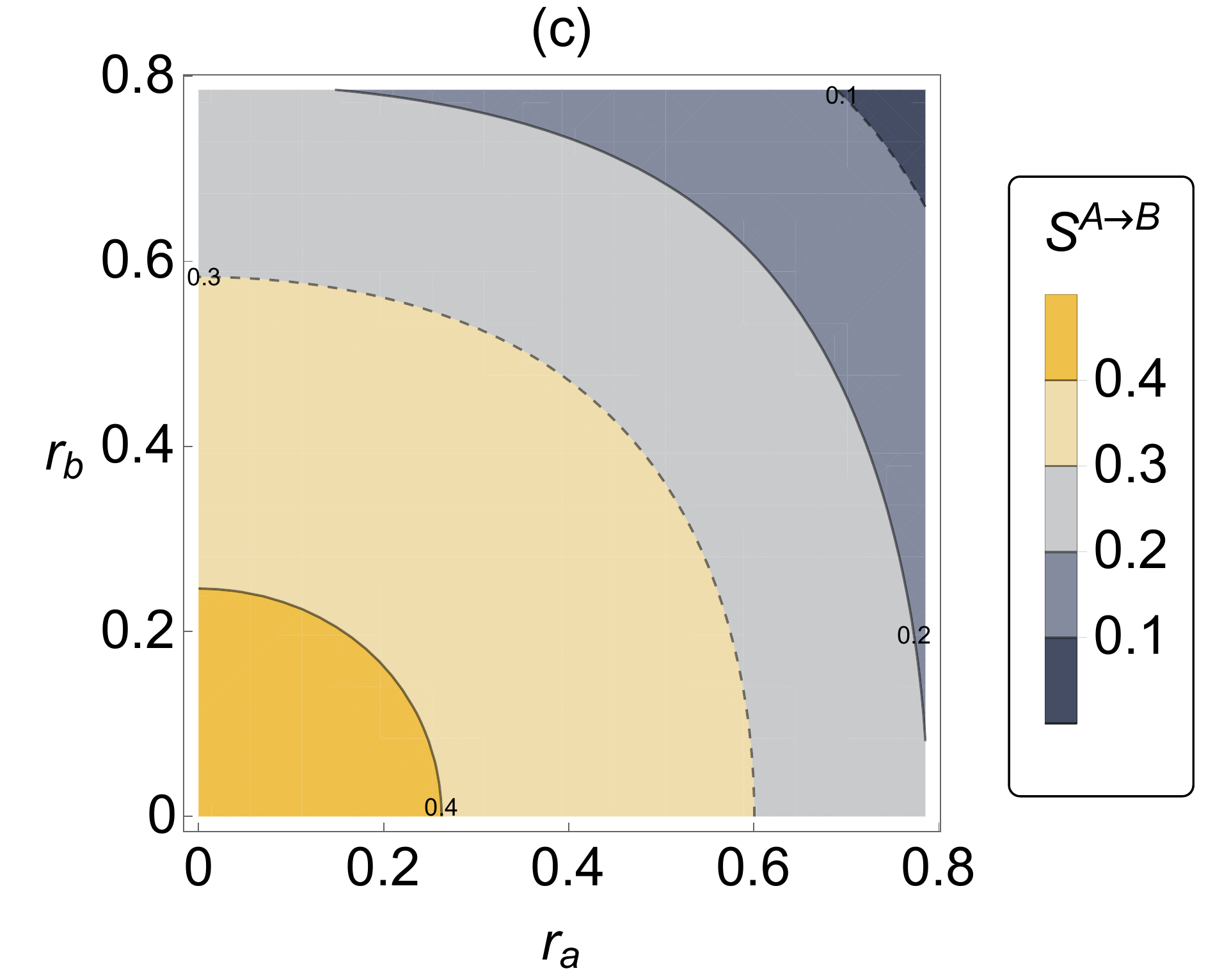}
	\caption{Steerability  from Alice to Bob by using the  generic pure state, when we set $p=0.5$, (a) $ \alpha=0.1 $ .(b) $ \alpha=0.4 $, (c) $ \alpha=0.7 $.}
	\label{fig:c3}
\end{figure}

The effect of the filtering process on an  initial state of a generic pure state  with  $p=0.5$ is displayed in Fig.(\ref{fig:c3}), where it is assumed that, both qubits of the shared state  are accelerated. In this situation, the decoherence  arises only from the acceleration process, and consequently the  degree of steerability decreases as the acceleration  of both partners increases. However, as one switches on the filtering process, the possibility of achieving the steering process  increases, namely the steering inequality is satisfied at large values of the acceleration.  This  phenomenon is displayed by comparing Figs.(\ref{fig:c3}a), (\ref{fig:c3}b), and (\ref{fig:c3}c), where the steerable areas are enlarged as the filtering parameter $\alpha$ increases. It is worth to mention that, the filtering process does not increase the degree of steerability, but increase the range of steerability.

\begin{figure}[h!]
	\centering
	\includegraphics[width=0.32\linewidth, height=4cm]{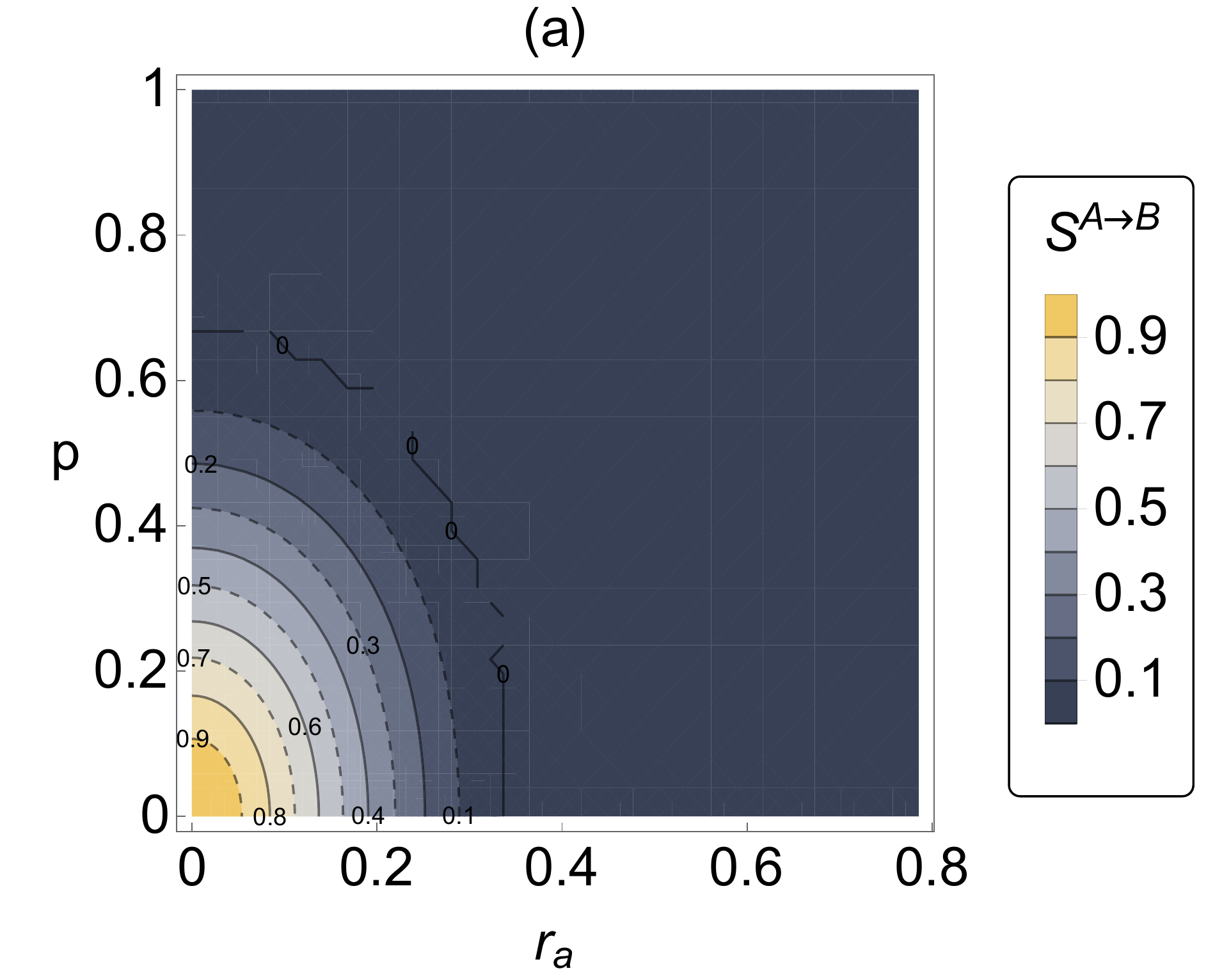}
	\includegraphics[width=0.32\linewidth, height=4cm]{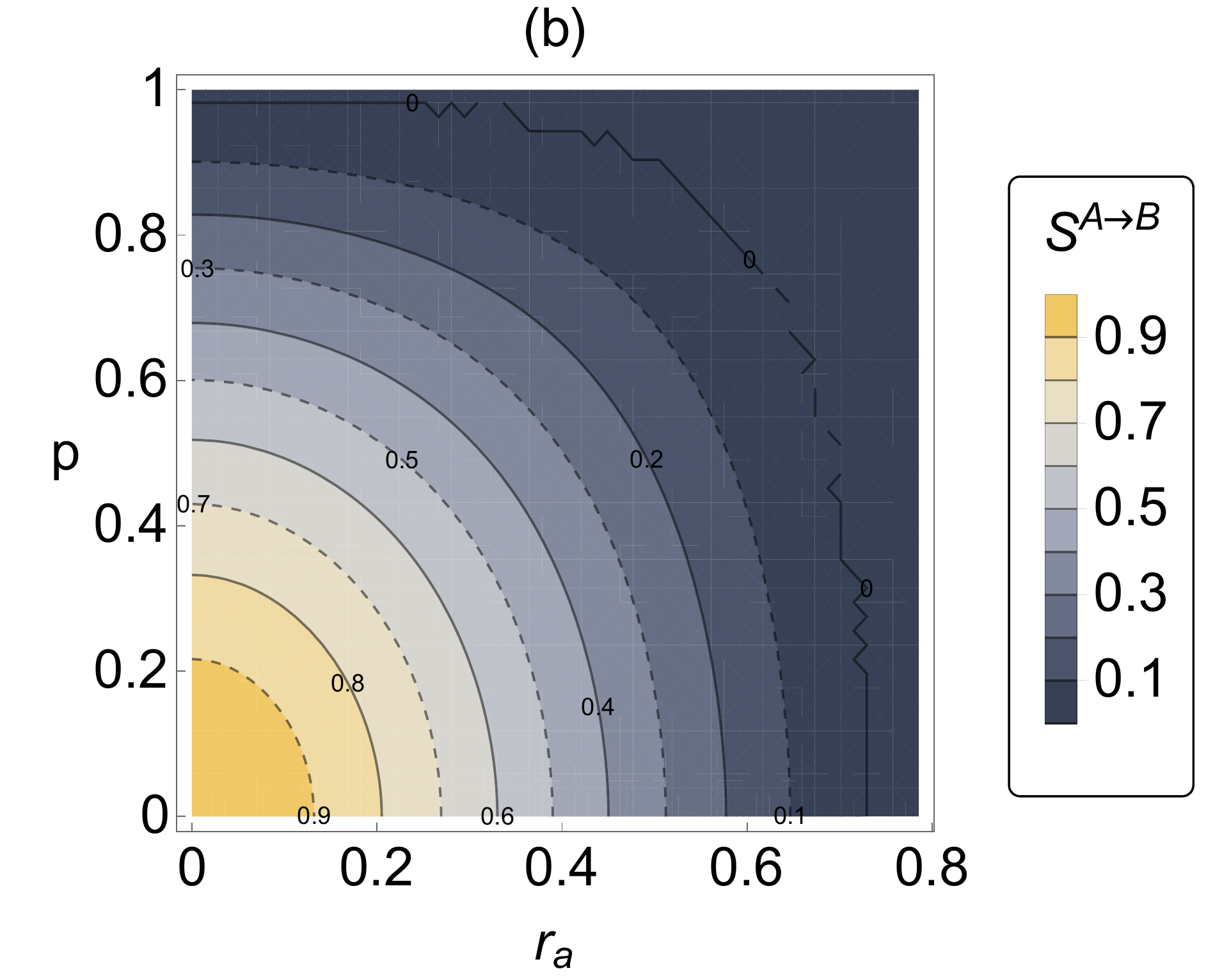}
	\includegraphics[width=0.32\linewidth, height=4cm]{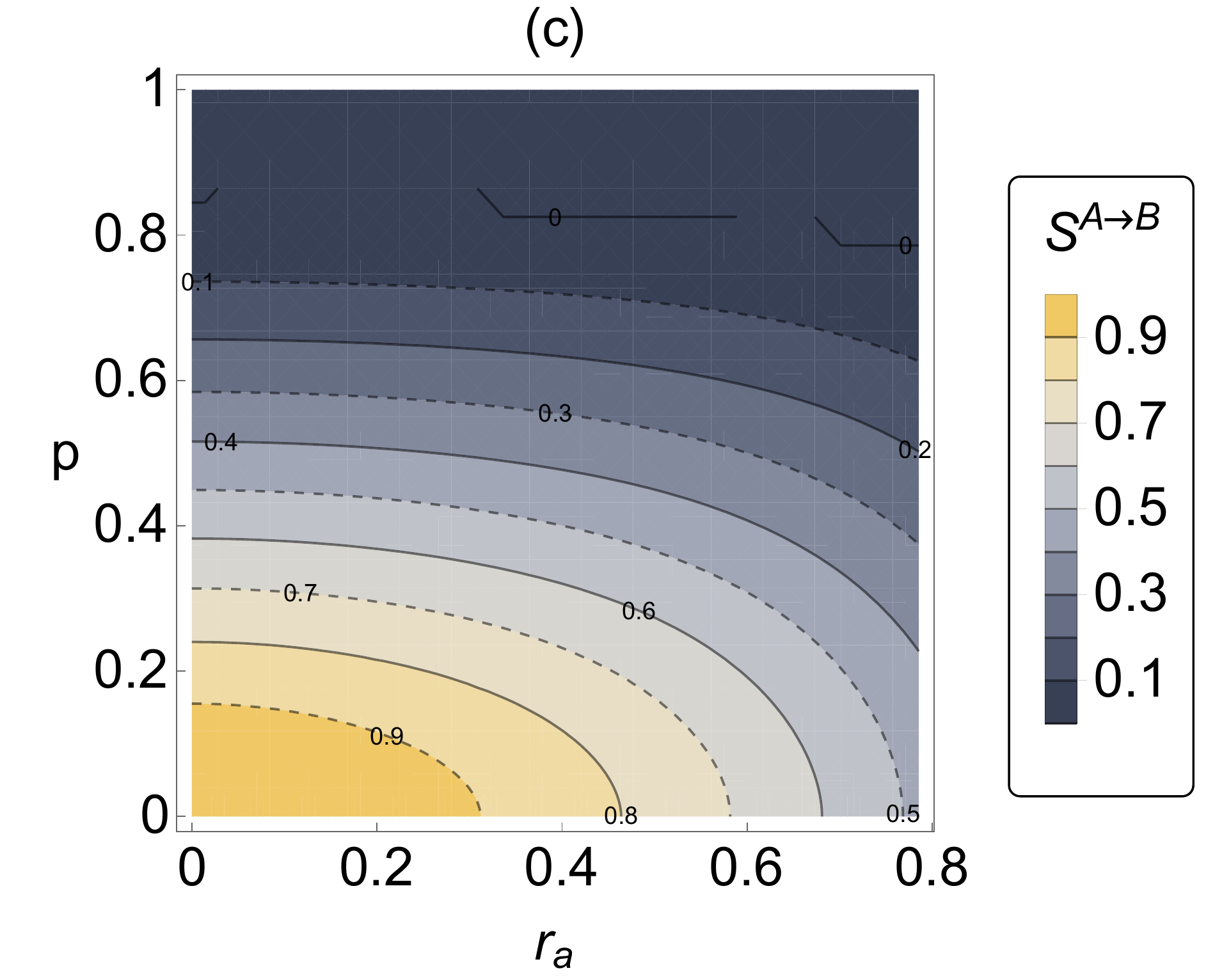}
	\caption{Steerability  from Alice to Bob for generic pure state, where Alice is accelerated with $ r_a $, and Bob at rest $ r_b=0 $. (a) $ \alpha=0.1 $ .(b) $ \alpha=0.4 $, (c) $ \alpha=0.8 $.}
	\label{fig:6}
\end{figure}
To  complete the discussion on the effect of the filtering process on the steerability and its degree, we consider  several classes of initial state settings, where we plot the function $\mathcal{S}^{A\longrightarrow  B}(p,r_a,r_b=0)$ in Fig.(\ref{fig:6}). In this case, there are two sources of decoherence: one due to the initial state settings, where the entanglement of the shared state decreases as $p$ increases, and the second due to the acceleration process.  Therefore, as it is expected, the degree of steerability decreases as one increases the decoherence  parameter $p$, and the acceleration $r_a$.  Figs.(\ref{fig:6}a-6c) show that, the filtering process has no effect on the upper bounds of $\mathcal{S}^{A\longrightarrow B}$. The comparison of these figures shows that, the range of steerability is enlarged  as the filtering parameter increases. This means that, the steering inequality is satisfied at larger range of the acceleration.

\section{Bidirectional steerability} \label{S3}
In this section, we show that for the  generic pure state the  degree of steerability from Alice to Bob and  vise versa, namely $\mathcal{S}^{B\longrightarrow A}$ depends on the type of steered information.
Fig(\ref{Bi-1}) shows the behavior of the bidirectional steerability when only Alice is accelerated and different initial state settings are considered.
The general behavior shows that, the degree of steerability decreases as the acceleration increases. However, as it is displayed in Fig.(\ref{Bi-1}a), if the partners share a maximum entangled state (singlet), both degrees of steerability coincide  at small values of the acceleration, i.e. $r_a\in[0,0.4]$. However, at  larger values of $r_a$, the degree of steerability from Bob to Alice, $\mathcal{S}^{B\longrightarrow A}$ is larger than that displayed from Alice to Bob, $\mathcal{S}^{A\longrightarrow B}$. On the other hand, since Alice particle is accelerated, then its decoherence increases and, consequently its local information decreases.  Therefore, one can say that the possibility of  steering  states that coded  coherent information  is much larger than that coded mixed information.
\begin{figure}[!h]
	\centering
	\includegraphics[width=0.32\linewidth, height=3.9cm]{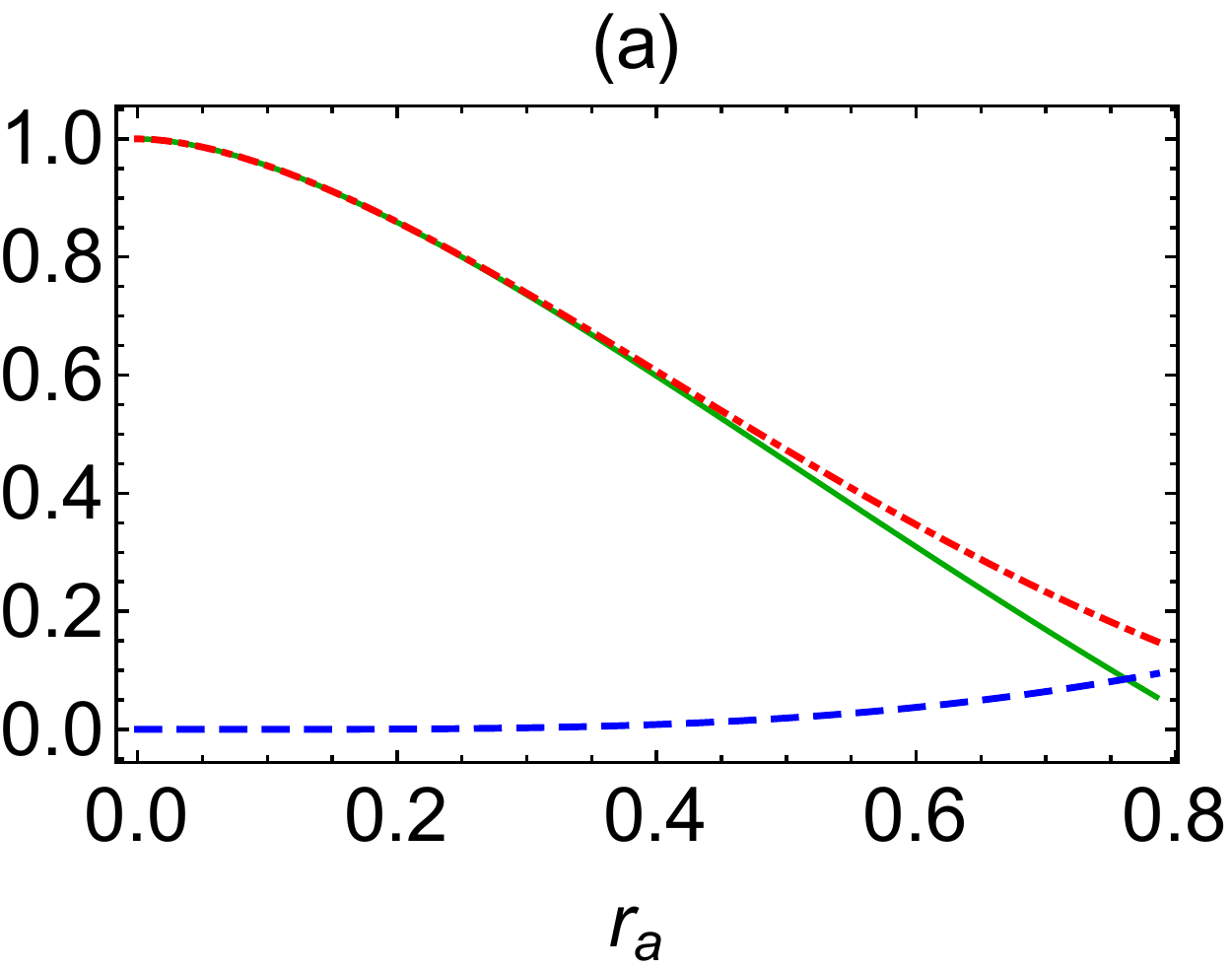}
	\includegraphics[width=0.32\linewidth, height=3.9cm]{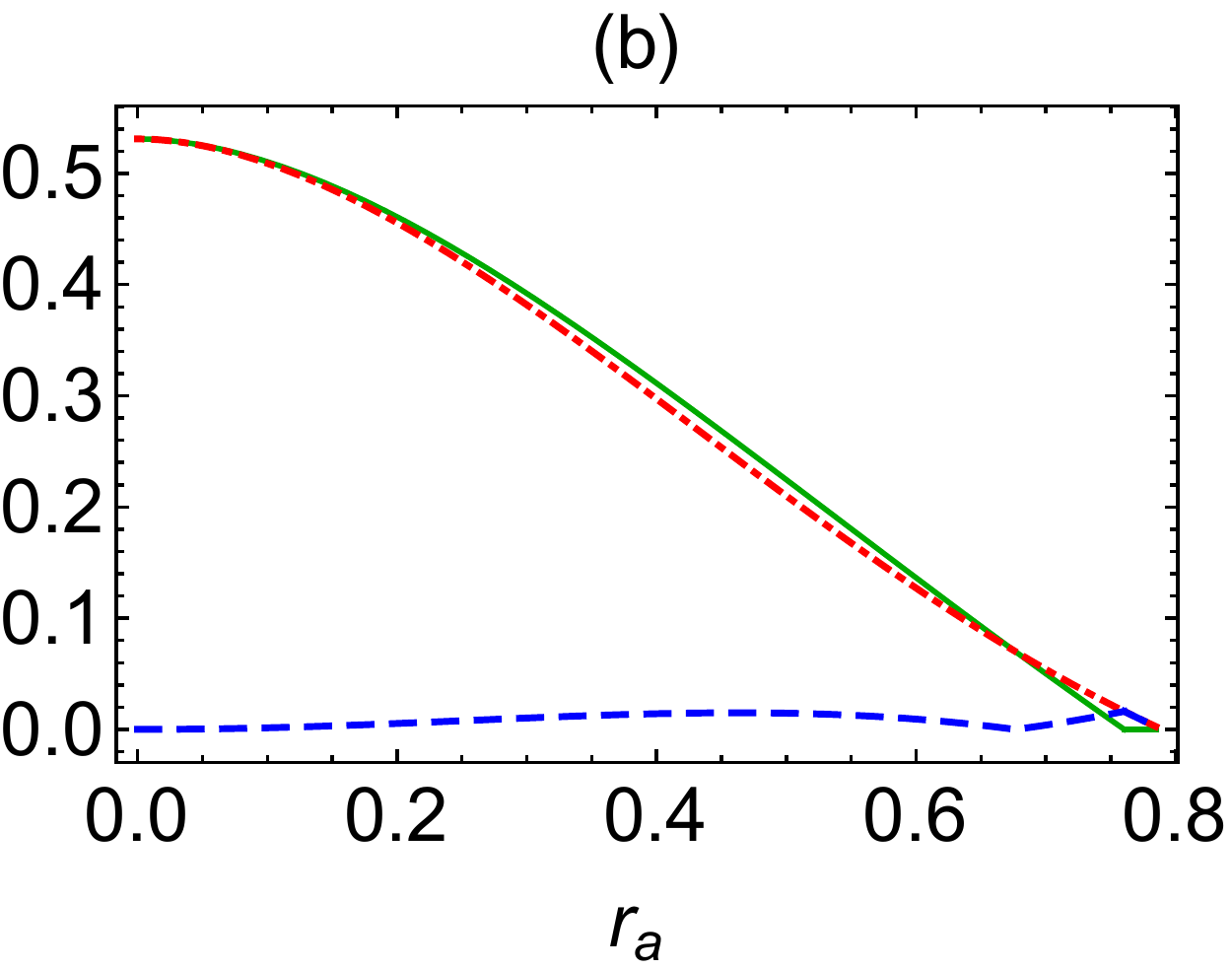}
	\includegraphics[width=0.32\linewidth, height=3.9cm]{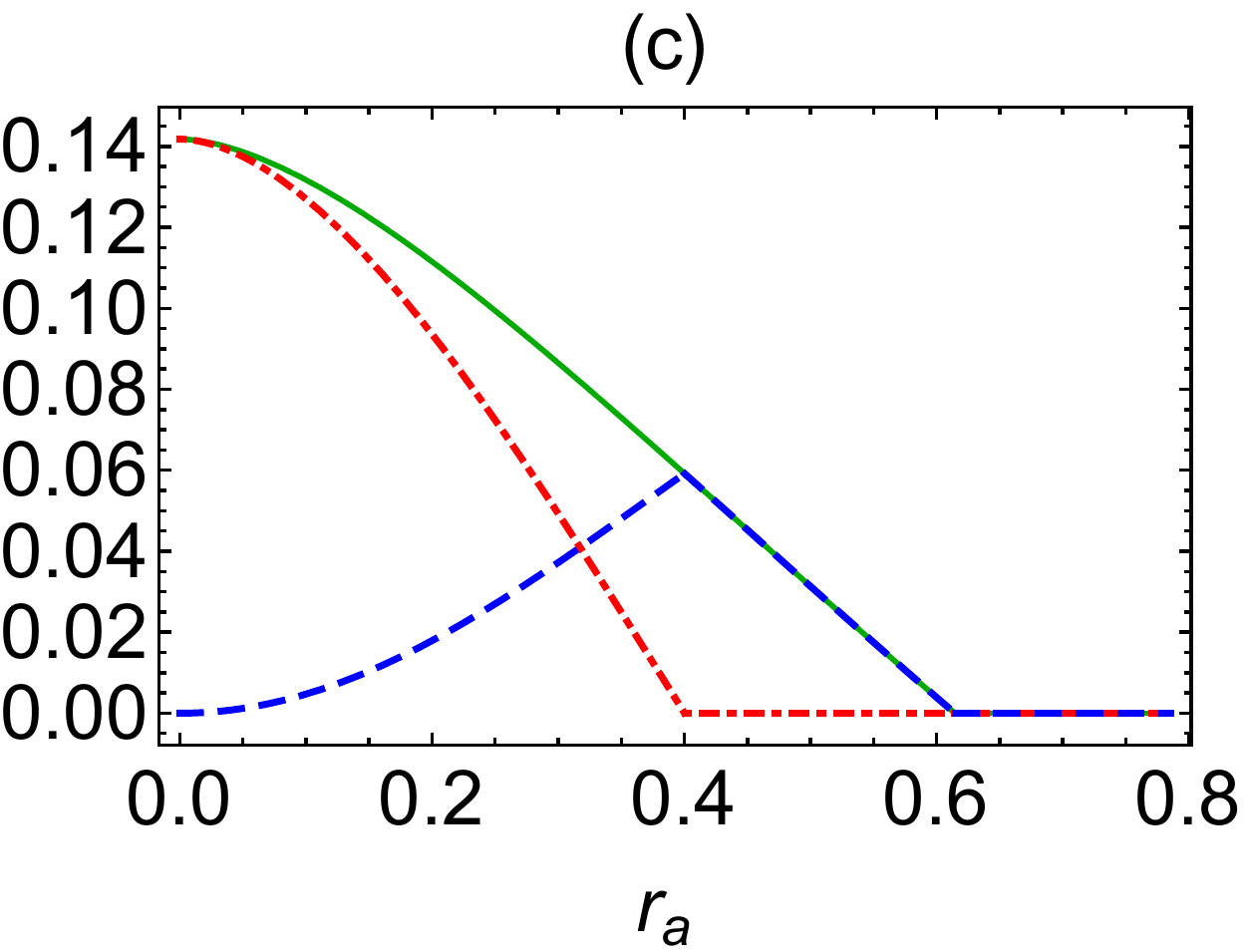}
	\caption{ Quantum steering for Alice is accelerated $r_a$, and Bob is fixed ($r_b=0$) where $\mathcal{S^{A\longrightarrow B}}$(green-solid curve),$\mathcal{S^{B\longrightarrow A}}$(red- dot dash curve), $\Delta=|\mathcal{S^{A\longrightarrow B}}-\mathcal{S^{A\longrightarrow B}}|$ (blue-dash curve). (a)$p\longrightarrow0$, (b)$p=0.6$, and (b)$p=0.9$ .}
\label{Bi-1}
\end{figure}

In Fig.(\ref{Bi-1}.b), it is  assumed that the partners  share a partially entangle state with $p=0.6$.  A similar behavior is depicted in Fig.(\ref{Bi-1}a), where the degree of steerability decreases as the  acceleration  $r_a$ increases. The displayed behavior shows that $ \mathcal{S}^{A\longrightarrow B}$ and $\mathcal{S}^{B\longrightarrow A}$ have a  slight  difference,such that $\mathcal{S}^{A\longrightarrow B}>\mathcal{S}^{B \longrightarrow A}$. This means that, the decoherence due to the acceleration is smaller than that due to the  decoherence parameter $p$.  Consequently, the  encoded  local information on Bob's qubit is more decoherent.  Fig.({\ref{Bi-1}c), shows the bidirectional steerability from both sides at larger value of the characteristic parameter $p$, namely the  shared state has small degree of entanglement.  The  behavior of the steerability $\mathcal{S}^{A\longrightarrow B}$, is larger than that displayed for $\mathcal{S}^{B\longrightarrow A}$ and the difference between them $\Delta$, increases as $r_a$ increases.  In the interval $r_a\in[0.1,0.8]$,  Bob's information is more coherent, and consequently the degree of Alice steerability $\mathcal{S}^{A\longrightarrow B}$ is larger than that displayed for  $\mathcal{S}^{B\longrightarrow A}$.  From Fig.(\ref{Bi-1}), it is clear that the  bidirectional degree of  steerability decreases as one increases the  parameter $p$.  The rate of decreasing due to the acceleration is much smaller than that displayed for the less entangled   shared state between  the steerer and the steered.
	
\begin{figure}[h!]
	\centering
	\includegraphics[width=0.32\linewidth, height=4cm]{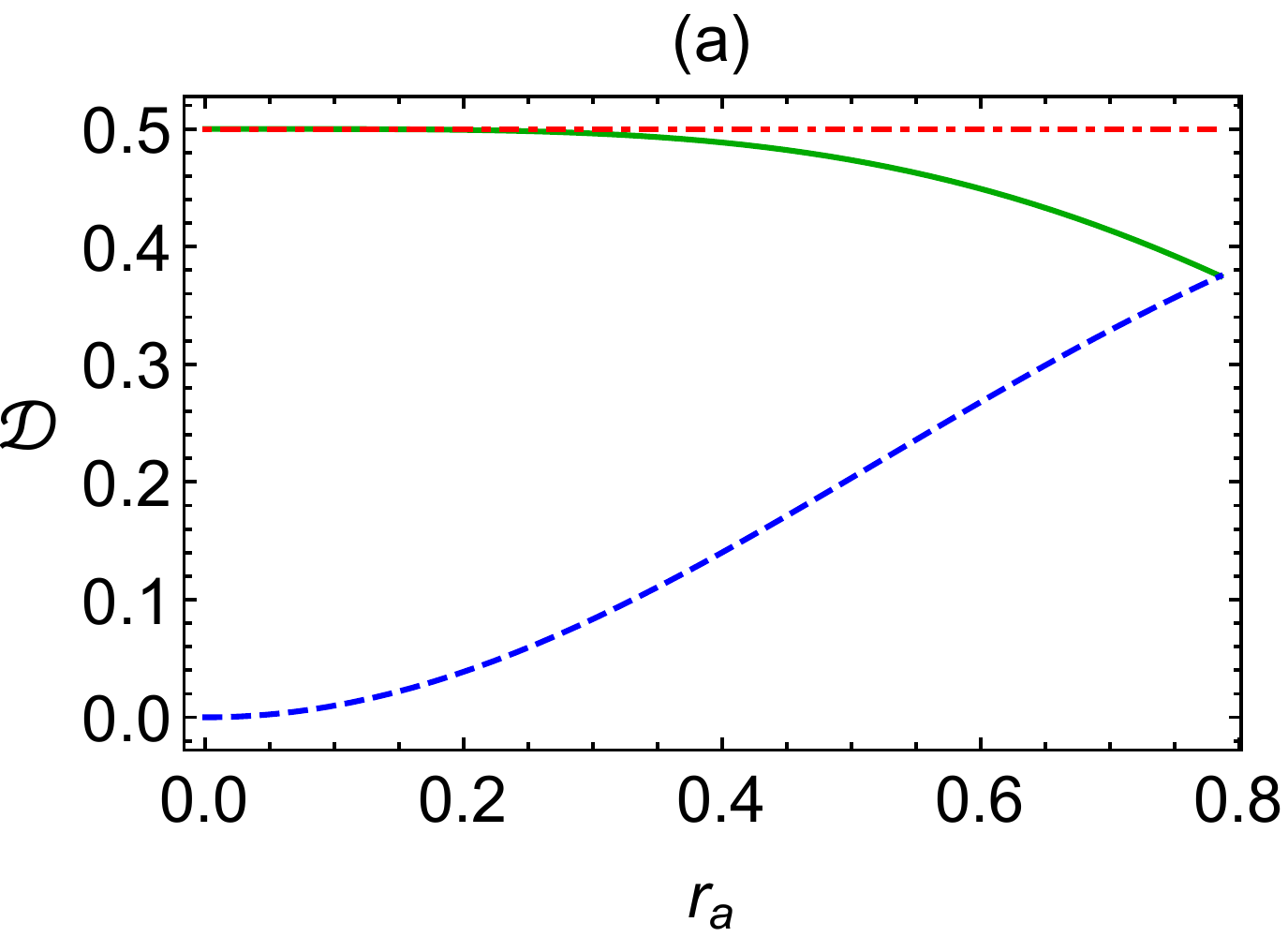}
	\includegraphics[width=0.32\linewidth, height=4cm]{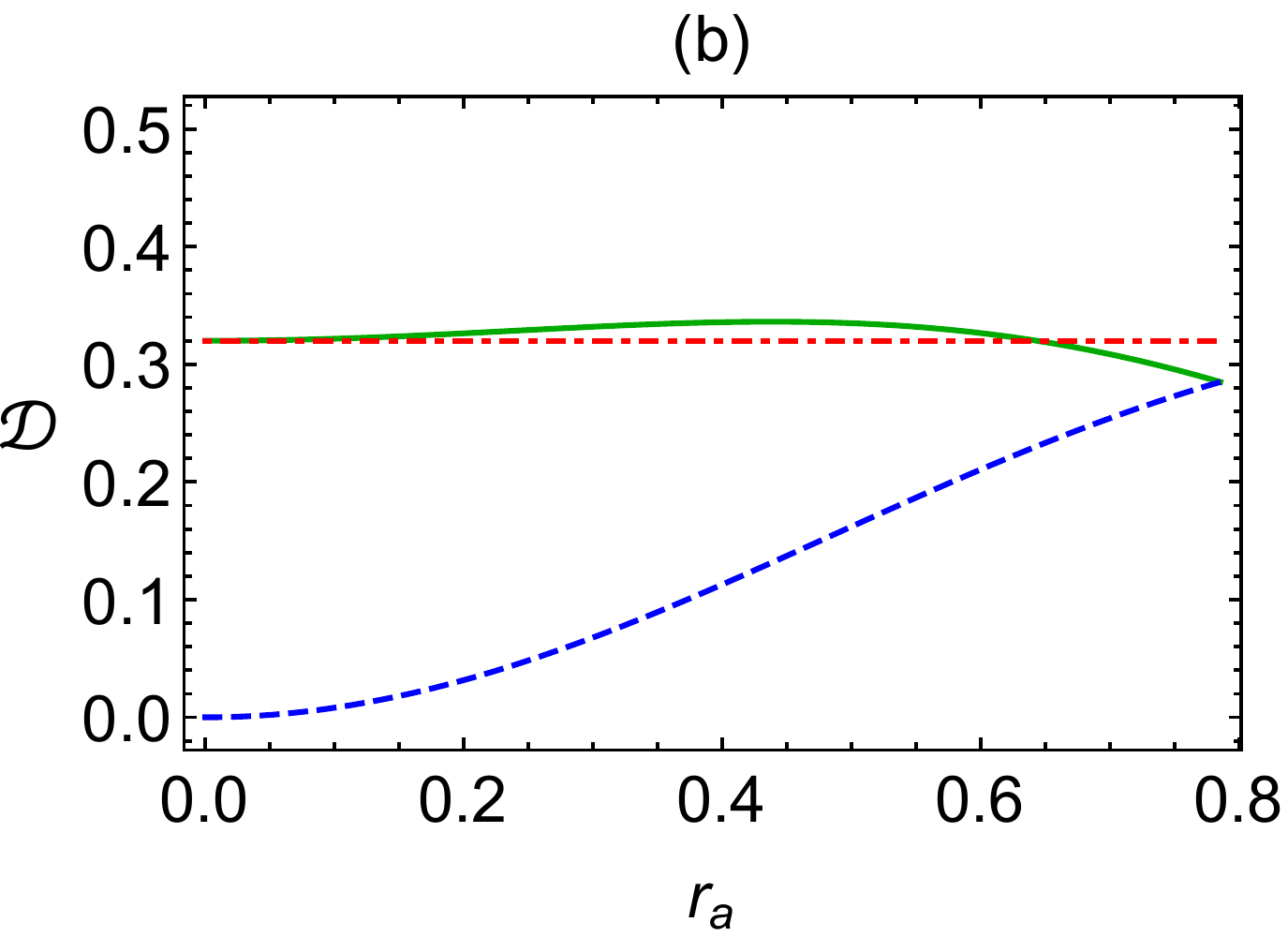}
	\includegraphics[width=0.32\linewidth, height=4cm]{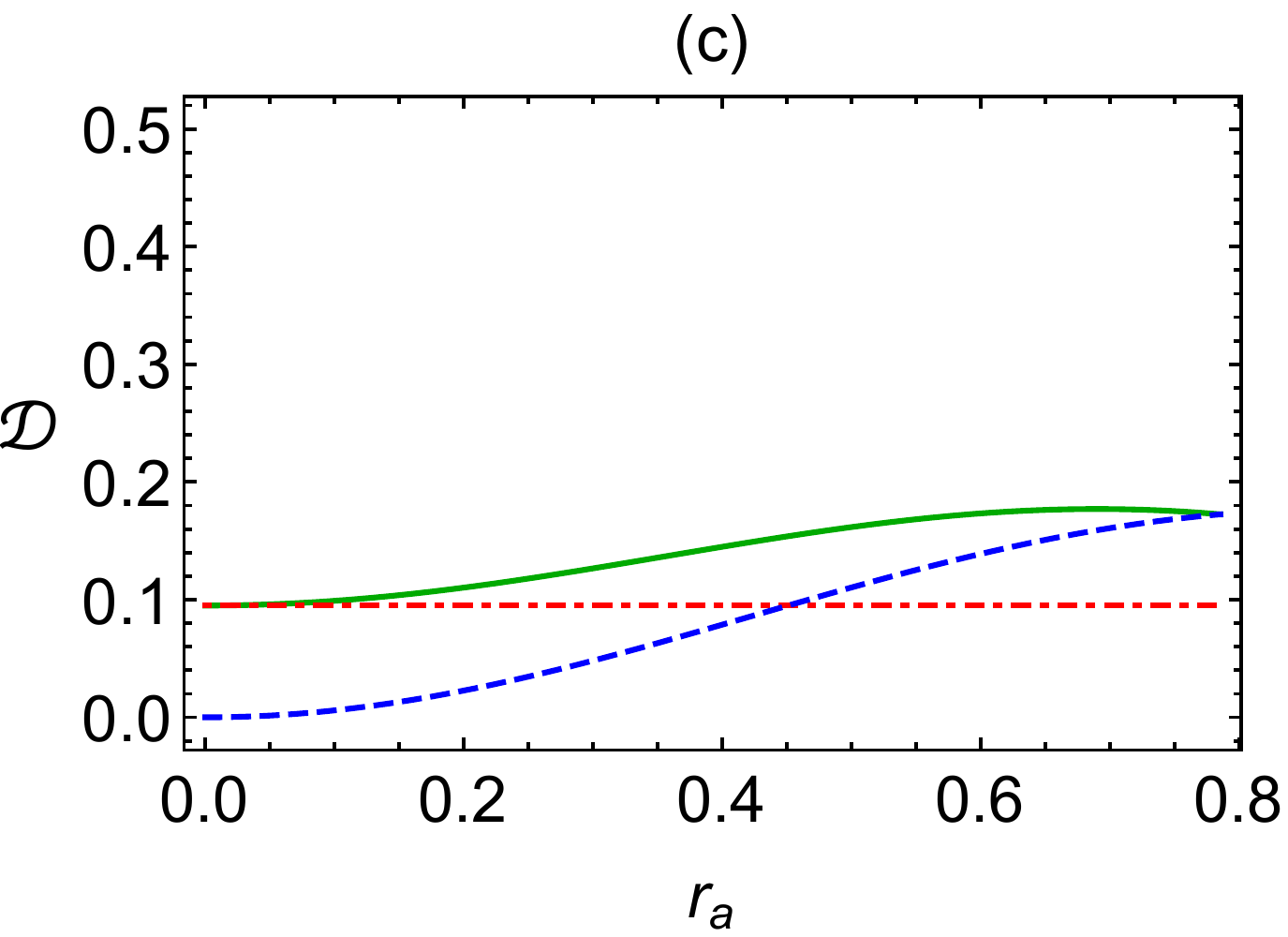}
	\caption{decoherence for generic pure state where Alice is accelerated,  decoherence $ \rho_{AB} $ (Blue curve), decoherence $ \rho_{A} $ (green curve),  decoherence $ \rho_{B} $(red curve), (a)$p=0.0$, (b)$p=0.6$, (c)$p=0.9$.}
	\label{fig:2}
\end{figure}

 \section{Conclusion}\label{s6}
  In this contribution, we investigated the bidirectional steerability between two partners who share different classes of initial states. It is assumed that, either the steerer or the steered partners are  in inertial/ non-inertial frames.  The effect of the initial state settings and the acceleration parameters on the steerability and its degree are  discussed, where analytical forms are obtained for both quantities. However, the possibility of enhancing the degree of steerability is discussed by applying the filtering process.

  The general behaviour of the steerability and its degree decreases,  if both  partners's qubits are accelerated.  The decay rate depends on whether one or both qubits are accelerated. It is shown that, the bidirectional steering is completely consistent if both partners are in the inertial frame or in the non-inertial frame with the same value of the acceleration.  The bidirectional steerability and its degree are slightly different when one of the partners's qubit is at rest and the other is accelerated.
  The initial state settings play a significant role on  maximizing/minimizing the steerability and its degree. It is shown that, the violation of the steering inequality appears as the shared state loses its quantum correlation. The largest upper bounds of the steerability and its degree are exhibited for the maximum entangled state.  The two quantities are investigated for a class of a generic pure state. The violation of the steerability is displayed at larger values of this  parameter, where the shared state turns into a separable state when one increases the decoherence parameter. However, due to the acceleration, the violation of steerability is depicted at small values of this  decoherence parameter.

   The possibility of improving the steerability and its degree is discussed by applying the filtering process. The results show that, as one increases the strength of the filtering process, the steerability increases. The filtering process restrains the decoherence that may  be  arisen from the accelerating  process, and consequently, one enlarges the range of the acceleration in which the steerability could be achieved. At fixed acceleration, the filtering process increases the upper bounds of the degree of steerability.

   The bidirectional steerability between the two users is discussed for the non-symmetric cases. It is clear that, if the users' qubits are accelerated, then the degree of steerability between the two users coincide. It is shown that, for non-symmetric cases, the degree of steerability depends on the types of steerable information. However, due to the acceleration the coherence of the qubit decreases and the possibility of its steering it decreases.  In this case, the difference between the steerability from both directions increases.

   Finally, one may conclude that, the decay of the  steerability and its degree is due to the decoherence from the initial state settings and the acceleration process.  The filtering process improves the upper bounds of the degree of steerability and enlarges the range of acceleration in which the steerability could be achieved. The most  information holds a great degree of steerability.

\end{document}